\documentclass[natbib=false,manuscript]{acmart}
\renewcommand\footnotetextcopyrightpermission[1]{} 
\setcopyright{none}
\usepackage{booktabs} 

\usepackage{blindtext}

\usepackage{xpatch}

\makeatletter
\xpatchcmd{\ps@firstpagestyle}{Manuscript submitted to ACM}{}{\typeout{First patch succeeded}}{\typeout{first patch failed}}
\xpatchcmd{\ps@standardpagestyle}{Manuscript submitted to ACM}{}{\typeout{Second patch succeeded}}{\typeout{Second patch failed}}    \@ACM@manuscriptfalse
\makeatother

\usepackage{paralist}

\usepackage{etex}
\usepackage[T1]{fontenc}
\usepackage{wrapfig}
\usepackage{blindtext}
\usepackage{harpoon}

\usepackage{graphicx,color}
\usepackage{amsmath}
\usepackage{amsfonts}
\usepackage{amssymb}
\usepackage{url} 
\usepackage{multicol}
\usepackage{multirow}
\usepackage{setspace} 
\usepackage{xspace}
\usepackage{listings}
\usepackage{ifthen}
\usepackage{verbatim}
\usepackage{float} 
\usepackage{stmaryrd} 
\usepackage{textcomp}
\usepackage{todos} 
\usepackage{fixltx2e} 
\usepackage{booktabs}
\usepackage[position=b]{subcaption}
\usepackage[normalem]{ulem}
\usepackage{tikz}
\usepackage{textcomp}
\usetikzlibrary{shapes,positioning}
\usetikzlibrary{calc}
\usepackage{adjustbox}
\usepackage{lipsum}
\usepackage[mathletters]{ucs}
\usepackage{wasysym}	
\usepackage{mdframed}

\usepackage[noend]{algorithmic} 
\usepackage{algorithm}
\usepackage{balance}

\newcommand{\etal}{\hbox{\emph{et al.}}\xspace}
\newcommand{\eg}{\hbox{\emph{e.g.}}\xspace}
\newcommand{\ie}{\hbox{\emph{i.e.}}\xspace}

\newfloat{Protocol}{thp}{lop}
\newfloat{Program}{thp}{lop}
\newfloat{Procedure}{thp}{lop}

\DeclareUnicodeCharacter{183}{\cdot}						
\DeclareUnicodeCharacter{931}{\ensuremath\Sigma}			
\DeclareUnicodeCharacter{9001}{\ensuremath\langle}			
\DeclareUnicodeCharacter{9002}{\ensuremath\rangle}			
\DeclareUnicodeCharacter{9608}{\ensuremath\blacksquare}		
\DeclareUnicodeCharacter{1013}{\in}							
\DeclareUnicodeCharacter{8213}{---}							

\DeclareCaptionLabelFormat{noparens}{Figure #2:} 

\usepackage[utf8x]{inputenc}
\usepackage[T1]{fontenc}
\usepackage{microtype}

\newcounter{RQCounter}

\newlength{\emstr}
\newcommand{\boldpara}[1]{%
 \smallskip%
 \par\noindent\textbf{\textit{#1}}\hspace{\emstr}
}%

\newcommand\restr[2]{{
  \left.\kern-\nulldelimiterspace 
  #1 
  \vphantom{\big|} 
  \right|_{#2} 
  }}
  
  \lstdefinestyle{codeSnippet}{ belowcaptionskip=1\baselineskip, breaklines=true,
	xleftmargin= 0.5cm, language=C++, showstringspaces=true,
	basicstyle=\scriptsize\ttfamily,
	keywordstyle=\bfseries\color{green!40!black},
	commentstyle=\color{orange!40!black}, numbers=left,
	numberstyle=\tiny\color{gray}, captionpos=b, escapeinside=@@, tabsize=4,
	breakatwhitespace=true, showlines=true columns=fullflexible,
	morekeywords={then} }

 \lstdefinestyle{codeSnippetLLVM}{ belowcaptionskip=1\baselineskip, breaklines=true,
	xleftmargin= 0.5cm, language=C++, showstringspaces=true,
	basicstyle=\scriptsize\ttfamily,
	keywordstyle=\bfseries\color{green!40!black},
	commentstyle=\color{orange!40!black}, numbers=left,
	numberstyle=\tiny\color{gray}, captionpos=b, tabsize=4,
	breakatwhitespace=true, showlines=true columns=fullflexible,
	morekeywords={then} }

\usepackage{eqparbox}
\usepackage{amssymb}
\usepackage{amsmath}
\usepackage[percent]{overpic}

\usepackage[authoryear]{natbib} 
\setcitestyle{aysep={}} 
\setcitestyle{square}

\mathchardef\uminus="2D

\setcopyright{none}

\newcommand{\Ourtool}{\textsc{Indexify}\xspace}
\newcommand{\Ourproject}{\textsc{Indexify}\xspace}
\newcommand{\type}{\ensuremath{\mathcal{T}}\xspace}

\begin{document}

\title{Indexing Operators to Extend the Reach of Symbolic Execution}

\author{Earl T. Barr}
\affiliation{%
  \institution{CREST, University College London}
  \streetaddress{ Malet Place}
  \city{London} 
  \postcode{WC1E 6BT}
  \country{UK}
}
\email{e.barr@ucl.ac.uk}

\author{David Clark}
\affiliation{%
  \institution{CREST, University College London}
  \streetaddress{Gower Street}
  \city{London} 
  \postcode{WC1E}
  \country{UK}
}
\email{david.clark@ucl.ac.uk}

\author{Mark Harman}
\affiliation{%
  \institution{CREST, University College London}
  \streetaddress{Gower Street}
  \city{London} 
  \postcode{WC1E}
  \country{UK}
}
\email{mark.harman@ucl.ac.uk}

\author{Alexandru Marginean}
\affiliation{%
  \institution{CREST, University College London}
  \streetaddress{Gower Street}
  \city{London} 
  \postcode{WC1E}
  \country{UK}
}
\email{alexandru.marginean.13@ucl.ac.uk}

\newtheorem{thrm} {Theorem}
\newtheorem{lm} {Lemma}
\newtheorem{prp}{Proposition}
\newtheorem{sketch}{Proof Sketch}

\begin{abstract}
Traditional program analysis analyses a program language, that is, all programs that can be written in the language. There is a
difference, however, between all possible programs that can be written and the corpus of actual
programs written in a language. We seek to exploit this difference: for a given program, we apply a bespoke program
transformation (\textsc{indexify}) to convert expressions that current SMT
solvers do not, in general, handle, such as constraints on strings, into equisatisfiable
expressions that they do handle.  To this end, \textsc{indexify} replaces
operators in hard-to-handle expressions with homomorphic versions that behave the same
on a finite subset of the domain of the original operator, 
and return $\bot$ denoting unknown outside of that subset.  By
focusing on what literals and expressions are \emph{most useful for analysing a
given program}, \textsc{indexify} constructs a small, finite theory that
extends the power of a solver on the expressions a target program builds.  

\textsc{Indexify}'s
bespoke nature necessarily means that its evaluation
must be experimental, resting on a demonstration of its effectiveness in practice.
We have developed \textsc{indexify}, a tool for \textsc{indexify} and released it
publicly. We demonstrate its utility and effectiveness by applying it to two
real world benchmarks --- string expressions in coreutils and floats in
fdlibm53.  \textsc{indexify} reduces time-to-completion on coreutils from Klee's
49.5m on average to 6.0m. It increases branch coverage on coreutils from
30.10\% for Klee and 14.79\% for Zesti to 66.83\%.   When indexifying floats in
\lstinline+fdlibm53+, \Ourtool increases branch coverage from 34.45\%
to 71.56\% over \lstinline+Klee+.  For a restricted class of inputs,
\textsc{indexify} permits the symbolic execution of program paths unreachable
with previous techniques: it covers more than twice as many
branches in coreutils as Klee.

\end{abstract}

%




\maketitle

\section{Introduction}

Symbolic execution (symex) supports reasoning about all states along a path.  It is
limited by its solver’s ability to resolve the constraints that occur in a
program’s execution.  Different types of symex handle intractable constraints differently. Broadly, static symbolic execution
abandons a path upon encountering an intractable constraint;  dynamic symbolic execution
concretises the variables occurring in the constraint and continues execution,
retaining the generality of symex only in the variables that
remain symbolic.  In this paper, we improve the responses of symex
for a specific program, allowing it to continue past intractable constraints
without resorting to fully concretising the variables involved in that
constraint.

Our approach is \Ourproject,
a general program transformation framework
that re-encodes intractable expressions, in any combination of types, into
tractable expressions. This is impossible in general but can be achieved to a
limited (and varying) degree for any particular program, so we characterise 
\Ourproject as program-centric. To transform a program, \Ourproject 
homomorphically maps a finite subset of the program's algebra of
expressions to an algebra of indices (or labels), augmented with the
undefined value $\bot$.  

\Ourproject is a program transformation that rewrites its input program to
replace operators with versions that 1) are restricted to $G$, a finite subset
of their original domain and range, and 2) take and return indices over $G$.
Let $\mathit{indexOf}$ map $G$ to $\mathbb{N}$.  When \Ourproject replaces the
operator $f$, its finite replacement $\hat{f}(\mathit{indexOf}(x)) =
\mathit{indexOf}(f(x)) \text{ if } x \in G$ and $\bot$ otherwise.  To build
$\hat{f}$, \Ourproject memoises the computation of $f$ over $G$.  To this end,
\Ourproject takes an input $P$, a (small) set of types $\mathcal{T}$, and
(small) set of literals, $S$, of type $\mathcal{T}$, as seeds.  In this work,
we harvest constants from the program as seeds, as we explain in
\autoref{sec:evalg}.  It takes two sets of operators $B$ and $F$, not
necessarily distinct, where $F$'s elements occur in $P$ and may produce
intractable constraints.  \Ourproject transforms a program $P$ in stages.
First, \Ourproject repeatedly and recursively applies the operators in $B$ to
expand $S$ to a larger set, $G$, the ``Garden''.  Then, it memoises $f \in F$
over $G$ to produce $\hat{f}$.  Finally, it rewrites $P$ to use the memoised
versions of $F$.

\begin{figure}
\centering
\includegraphics[width=0.85\textwidth]{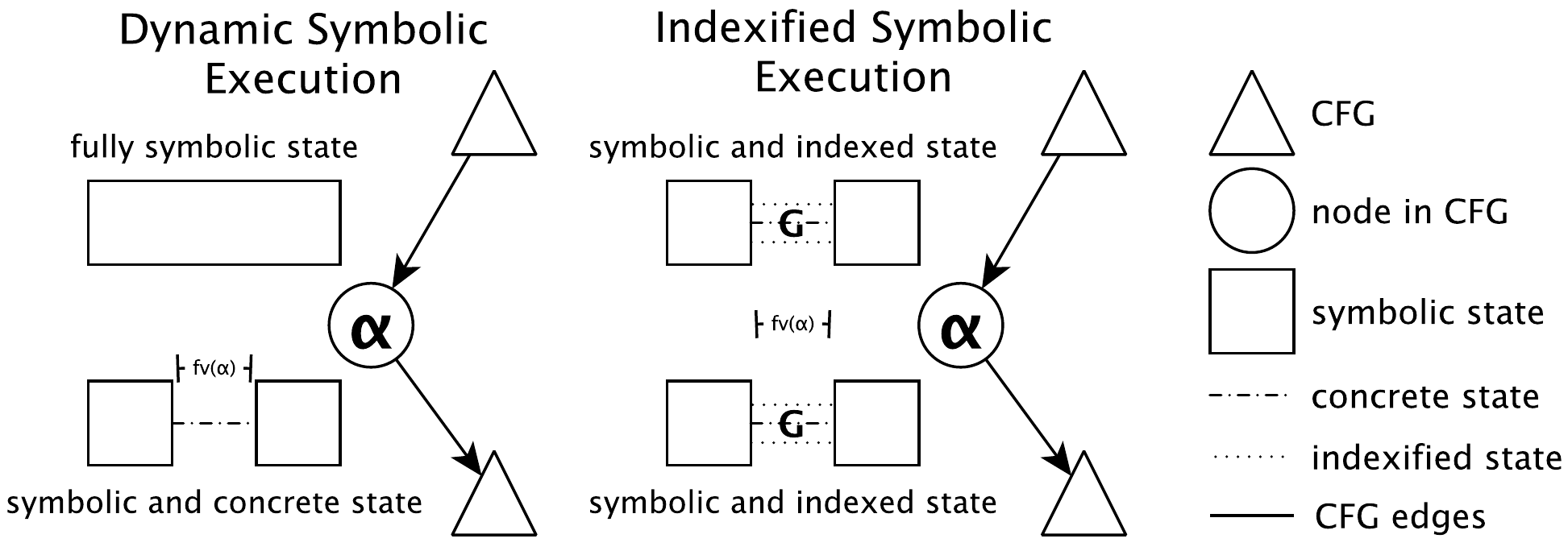}	

  \caption{Standard dynamic symex compared to indexified symex; the control
  flow graph (CFG) node is a control point and the solver returns unknown on
  queries containing $\alpha$.} 

\label{fig:index}
\end{figure}

Consider \autoref{fig:index}, which depicts dynamic symbolic execution (DSE).
In the figure, $\alpha$ is intractable, so the solver returns unknown when
queried about $\alpha$. On the left, DSE replaces the free variables in
$\alpha$ with concrete values, either drawn from a concrete execution that
reaches $\alpha$ (concolic~\citep{marinescu2012make}) or generated using heuristics~\citep{chen2013state}, collapsing the state space of $\alpha$'s variables to those
concrete values, but allowing symex to proceed over the rest of the state
space.

On the right, we see \Ourproject in action on a transformed program $P$.  Over
$\alpha$'s free variables, the program is restricted to the garden $G$.  An
indexed expression, like $\alpha$, that contains $\hat{f}$ evaluates to $\bot$
when it takes as an argument either $\bot$ or an unindexed value or it
evaluates to an unindexed value.  Indexed operators like $\hat{f}$ are lookup
tables that generate constraints in equality theory as we show in
\autoref{sec:constraint}.  \Ourproject's transformation permits symbolic
reasoning over the algebra of indices at the cost of explicit ignorance,
reified in $\bot$, about values outside the index set; it allows the symbolic
exeuction of the the indexed program over a (previously intractable) subset of
the original program's state space, permitting the symbolic exploration of
previously unreachable regions of that state space.  If we indexify
the problematic operators in $\alpha$, symex can continue,
restricted to $G$, past $\alpha$ in \autoref{fig:index}.  Once we apply
\Ourproject to a program, the problem becomes identifying a useful set of
expressions whose indexification might improve our ability to find bugs.

An indexed program under-approximates the semantics of the original program in
the following way:  if a transformed program's execution stays entirely within
the indexed subset of values and operators, its output is either an
index that is an image under the homomorphism of the original program's output
or it is undefined (returns $\bot$).  The under-approximation of the semantics
and its dependency on the choice of the algebra to indexify again emphasises
the bespoke nature of the transformation and its dependence on the goodness of
the choice. As we demonstrate later through examples, in practice, it is not
difficult to make a good choice.  Our main contributions follow
\begin{compactenum} 
\item The introduction and formalisation of a general framework for 
restricting operators to produce tractable constraints, allowing symbolic
  execution to explore some paths previously only concretely reachable 
  (\autoref{sec:formalism});
\item The realisation of our framework in the tool \Ourtool (\autoref{sec:imp}) available at <url>;
and

\item Comprehensive demonstrations of Indexify’s utility (\autoref{sec:res}):
  we compare \Ourtool with dynamic symbolic execution, \ie
  Klee~\citep{cadar2008klee}, and concolic testing, \ie 
  Zesti~\citep{marinescu2012make}. We show that \Ourtool achieves 66.83\%
  branch coverage, compared to Klee's 30.10\% and Zesti's 14.79\%, on
  coreutils, and does this within less than a third of the time that Klee
  without \Ourtool requires on average. Finally, we show it reaches bugs that Klee alone
  does not on a famous C bug finding benchmark.

\end{compactenum}

\section{Motivating Example}
\label{sec:example}

\begin{figure}[t]
\centering
\begin{lstlisting}[style=codeSnippet, firstnumber=47, xleftmargin=6mm]
+ typedef enum {NVidia, NVidiaCorporation, ...} string_enum;
- static const char * Nv11Vendor = "NVidia Corporation";
+ static const string_enum Nv11Vendor = vendor11;

BOOLEAN IsVesaBiosOk(...){ 
   ...
-  if (!(strncmp(Vendor, Nv11Vendor, sizeof(Nv11Vendor))))
+  if (!(i_strncmp(Vendor, Nv11Vendor, sizeof(Nv11Vendor))))
-    assert(strncmp(Vendor, Nv11Vendor, MAXLEN) == 0);
+    assert( i_strncmp{(Vendor, Nv11Vendor, MAXLEN) == 0);
}
\end{lstlisting}
\caption{Diff of buggy code from ReactOS, after \Ourtool: the assertion, which we added  to reify the bug, at line 55/56 can fail;  this bug is difficult to reach under either pure or concolic symbolic execution;  {\sc Indexify} reaches and triggers the bug by restricting variables to a finite set of indices (adding line 47) and by replacing types, constants (line 48/49), and operators (lines 53/54, and 55/56) to work with these indices.} 
\label{lst:example}
\end{figure}

\begin{figure}[t]
\centering
\begin{lstlisting}[style=codeSnippet,numbers=none]
Garden: 	@$G=\{ \emptyset, N, V, i, d, ... , NV, aN, ... \}$@
IOT:		int i_strncmp(const string_enum lhs, const string_enum rhs, int count){
				if(lhs == vendor && rhs == vendor11 && count == 4 return 0;	
				else if(lhs == vendor && rhs == vendor11 && count == 8) return 1;
				... return -1; // -1 represents @$\bot$@ 
			}
\end{lstlisting}
\caption{The garden, and the indexed operator tables IOTs  (indexed initial operators) that \Ourproject generates for the code snippet in \autoref{lst:example}. We use the Kleene Closure to build this example: $G$ is the Kleene Closure of all the strings in ReactOS. We build the IOT \lstinline+i_strncmp+ for the operator \lstinline+strncmp+.
\autoref{sec:imp} describes how \Ourtool achieves the above.\looseness=-1}
\label{lst:garden:iot}
\end{figure}

\autoref{lst:example} presents a code fragment that contains a real world bug
in \lstinline+ReactOS+~\citep{reactOSWeb},  an open-source
operating system. Commit \lstinline+5926581+, on 21.12.2010,  changes the type of variable \lstinline+Nv11Vendor+ from array to pointer. This causes the \lstinline+sizeof+ operator to return the size of a pointer,  not the length of the array. ReactOS developers fixed this bug in commit \lstinline+30818df+, on 18.03.2012, after ReactOS's developer mailing list\footnote{\url{https://reactos.org/archives/public/ros-dev/2012-March/015516.html}} discussed it. The developers fixed the bug one year and three months after it was committed.\looseness=-1

\autoref{lst:example} shows only the relevant code, after using \Ourtool,  in diff format.  Although \Ourproject operates at \lstinline+LLVM+ bitcode level, we use source code here for
clarity.  \Ourproject adds line 47 to define indices as an
\lstinline+enum+. \autoref{lst:garden:iot} shows the garden, and the indexed version of the operator \lstinline+strncmp+ (\lstinline+i_strncmp+). More details about the garden, the indexed operators, and how we build them, are in \autoref{sec:imp}.\looseness=-1

The bug is a classic error: line 57 (lines 53/54 in \autoref{lst:example}) in \lstinline+vbe.c+ of ReactOS, commit \lstinline+30818df+,  incorrectly applies
\lstinline+sizeof()+ to a string pointer, not
\lstinline+strlen()+~\citep{wagner2000first}.  As a result,
\lstinline+strncmp()+ compares only the first 4 characters of its 
operands, assuming a 32 bit pointer.  
A pair of strings, whose first four characters are identical then differ afterwards in at least one character, triggers the bug when bound to
`\lstinline+Vendor+' and `\lstinline+Nv11Vendor+': the
\lstinline+if+ condition on lines 53/54 wrongly evaluates to true and
assertion on lines 55/56 fails.  This bug is an error in a string expression and the theory of strings is undecideable, when the string length is unbounded~{\citep{quine1946concatenation}. Thus, most string solvers return UNKNOWN on this constraint~\citep{bjorner2009path}.

Static symbolic execution must content with intractable constraints. CBMC, for instance, errors on them~\citep{kroening2014cbmc} and would not reach the assertion on line 55.
Klee~\citep{cadar2008klee} implements dynamic symbolic execution and uses bit-blasting. When  Klee reaches the \lstinline+if+ on line 53, 
Klee internalizes \lstinline+strncmp+ to bitblast it.  The  \lstinline+strncmp+ function loops over the length of strings,  causing Klee's default solver to time out with its default settings ($1$ minute time-out). 
 Thus, Klee does not produce an input that triggers this bug.   Concolic testing searches a neighbourhood
around the path executed under its concrete inputs. Upon reaching exit,
concolic testing backtracks to the nearest condition, complements it, then
restarts execution from the entry point~\citep{godefroid:dart, sen2005cute}. Thus, concolic testing can reach this
bug only if  it is given a concrete input in the neighbourhood of this bug. Unlike concolic testing, which is tethered to
a single concrete execution, \Ourproject can symbolically reason over all the values in its finite set $G$, broadening its exploration relative to 
concolic testing.

\Ourtool finitizes operators in undecideable theories by transforming them into finite lookup tables over values of interest thereby converting potentially undecideable expressions into decideable ones.
To
build $G$, the set of interesting values, \Ourproject harvests the constants in a program, such as the ones in
\autoref{lst:example}, as seeds, then concretely and repeatedly applies operators, such as Kleene
closure~\citep{kleene1951representation}, to the constants, up to a bounded number, to expand the set.  From the constants in
\autoref{lst:example}, this process produces \lstinline+NVidia+ and
\lstinline+NVidiaCorporation+ among others.  The lookup table for \lstinline+strncmp+ memoize the concrete results of repeatedly 
evaluating it on pairs drown from $G$.  In \autoref{lst:example}, to index the strings \Ourproject
introduces the \lstinline+enum+ on line 47, then, on lines 48--49, it changes
\lstinline+Nv11Vendor+'s type to int and replaces the constant to which it is initialized to
one of the indices introduced on line 47.  \Ourproject then replaces
 \lstinline+strncmp+ with \lstinline+i_strncmp+ on lines 53/54 and 55/56.
Indexing these and building \lstinline+i_strncomp+, the memoised looking table for \lstinline+strncmp+ over the values in $G$ is
sufficient to violate the assertion at line 55/56.\looseness=-1

\autoref{lst:garden:iot} shows the garden $G$: the extended set of constants that we consider in our analysis; and the indexed operator table ($IOT$) \lstinline+i_strncmp+, the memoization table for the operator \lstinline+strncmp+. We discuss these concepts in details in \autoref{sec:formalism} and in \autoref{sec:imp}. \lstinline+i_strncmp+ contains entries for all the values in $G$. If \lstinline+i_strncmp+ gets parameters that are out of $G$ we abandon the path and return $\bot$. This means that the values that flowed into \lstinline+i_strncmp+ are out of our analysis.\looseness=-1

Klee times out on the \lstinline+strncmp+ operator. When  Klee runs on the indexified version of the code in \autoref{lst:example} it successfully executes the indexed \lstinline+strncmp+ operator: \lstinline+i_strncmp+; and produces a bug triggering input. The bug-triggering constraint that \Ourtool generates is: $\text{Vendor} = 0 \land \text{Nv11Vendor} = 1$. The solver produces the values: $\text{Vendor} = 0$ and $\text{Nv11Vendor} = 1$. Under these inputs, \lstinline+i_strncmp+ returns $0$ signalling that the strings are equal. The assertion on line 56 fails, as the strings are equal only in the first 6 character but differ afterwards.

\section{Approach}
\label{sec:formalism}

The concept of \Ourproject is quite general. It aims to transform a program so as to
generate tractable constraints at \emph{some} points at which  it had previously
generated intractable or undecidable constraints  by restricting these
constraints to a simpler theory over a finite set of values, augmented with 
the unknown value, $\bot$.  The resulting, transformed 
constraints should be satisfiable whenever the original constraints were satisfiable, be more tractable with regard to satisfaction, and sometimes be satisfiable when the original constraints were not; but only when restricted to the finite set of chosen 
values and operators; crucially, SMT solvers can handle them efficiently.  This simpler
theory turns out in practice to depend on the way in which \Ourproject is
implemented, although in each case the overall approach is the same. 

To reiterate: for a given type or set of types we identify a useful set of literals and a desired set of operators on the literals, then memoize the outcomes of all combinations of applications of the operators on the literals --- but only up to a limit, $k$, on the number of applications in any one expression. This can be represented simply as a finite set of index tables, one for each operator of interest. Since the useful set of literals is not necessarily closed under applications of the desired set of operators,  we need to enter \emph{undefined} for some entries in the tables. The final step is to perform a program transformation by replacing all the members of the memoized sets of literals and operators, as they occur in the program syntax, with their indexified versions. The effect from the point of view of constraint solving 
 is that of shifting between logical theories. Solving constraints containing indexified operators can, as a result, use a more tractable theory such as equality.

This approach can be applied to any source theory, but, for the purposes of presentation and examples in the present paper, we restrict ourselves to  \lstinline+string+
expressions and their operators.  All non-trivial fragments of theory of strings are
NP-complete~\citep{jha2009computational}, and thus, string constraints are
intractable, making \Ourproject highly useful.

\subsection{Terminology}
\label{sec:terminology}

Logical theories can be viewed as algebras. In universal algebra, an algebra is an algebraic structure, that is, a set of literals and a set of operators on the literals, together with a set of axioms that collectively play the role of laws for the algebra. Sometimes the notion of an algebraic structure is simplified to just a set of operators and the literals appear as nullary operators. In what follows, we explain in detail the soundness requirement for the transformation in the program syntax, i.e. that it must be a partial homomorphism between two algebraic structures. This does not map logical laws between the algebras. In our setting, there is not necessarily a homomorphism for the logical laws; to see why consider that laws of the naturals and strings.

 To elaborate, in order to be sound, we require that the result of applying a transformed operator to transformed arguments yields the same result as applying the original operator to the original arguments and then transforming the result, whenever the result is defined in the transformed program. This is the minimum guarantee we should expect. Without it we could (unsuccessfully) transform any type to any type, any operator to any operator, e.g. strings to integers and replace operations on strings with operations on integers arbitrarily. In this section, we specify the behaviour of the homomorphism on the operators, then show that the algorithm for the transformation  constructs a homomorphism of this kind. Finally, the transformation is implemented as a rewrite system on the syntax of the program. 

It would be useful to be able to show that, whenever a transformed constraint has a model in the target theory, the untransformed version either has a model (but not necessarily the same model) in the source theory or is not satisfiable in the source theory. A proof of this would rely on properties of the individual theories and is left outside of the scope of this paper. Intuitively, given the homomorphism and given that the target theory is generally much simpler than the source theory, we believe that this will in general hold.  

Considering the set of literal values and the set of operators on them that may occur in a program, we can partition each set into those that we indexify and those that we do not. Those that we do not indexify are left untouched by the transformation and on these the homomorphism is simply the identity. For simplicity, we require that there is no interaction between the untouched parts and the transformed values and operators. In consequence, once we identify a set of operators to indexify, we must also indexify a ``sufficiently large set of values'' which are of the input and output types for this chosen set of operators. We could make other choices with regard to the relationship between the indexified and unindexified operators and values but this is the simplest choice.

\boldpara{Constructing a ``Garden" of Literal Values to Indexify:} Here,  we  formallly present how we target a set of type literals and a set of operators to expand the initial set (seeds) into a larger set of literals $G$ (garden). 

We begin our description with some useful
notation for types and operators. Let $X$ be a set of operators. Denote the subset of nullary operators
(literals) of $X$ as $X_0$ and the set of non-nullary operators by $X_+ =
X-X_0$. We will henceforth use the subscripts $0$ and $+$ to indicate sets of literals and sets of non-nullary operators respectively. Let $H$ be a function that takes a set and returns a set of the same type. Use $H^k(X)$ to mean the recursive application of $H$ $k$ times 
to the set $X$, so that $H^0(X) = X$ and $H^k(X) = H^{k-1}(H(X))$.

Suppose we have a program $P$ and want to indexify some of the operators that
occur in $P$.  Let $T$ be the set of types in $P$  and
$\oplus$ be the set of operators used in $P$.  Initially we have in mind a set of (non-nullary) operators of interest,  ones out of which are presumably potentially problematic for SMT solvers.  We first select a set, $S = B_0$,
of nullary operators (literals) whose types are basic types that include all the argument and return types of these operators of interest. We call this set the seeds. Then we select $B_+$, a set of non-nullary operators on these seeds that we use to build a larger set of
literals, the garden $G$. $B_+$ is not restricted to $\oplus$ and does not necessarily contain any of the operators of interest.  Each literal in $S=B_0$ has type
$
 \type, \text{where } \type= \tau_1 \uplus \tau_2 \uplus \cdots \uplus \tau_n
$.
In other words, $\type$ is a disjoint union of the types of nullary operators and each literal has one of those types.\looseness=-1

Let 
$f \in B_+ \Rightarrow f : \type \rightarrow \ldots \rightarrow \type \rightarrow \type$, \ie the argument and return types of $f$ are in $\type$. 

We define a function, $H: \type \rightarrow \type$ on a set of nullary operators, $Z$, as follows.
\[H(Z) = Z \cup \{f(x_1, x_2, \ldots, x_n) \mid f \in Z_+, x_i \in Z_0, \hat{f}(x_1, x_2, \ldots, x_n) \mbox{ is defined}\} \neq \bot \]
For simplicity of presentation, we have ignored all the ``side information'' about elements of $Z$ as arguments to $H$ in the type of $H$ so as to focus on its application as an iterative step in growing the garden. We can then define $G$, the garden resulting from $k$ applications of $H$, as 
\begin{align}\label{eq:buildG}
G = G_k = H^k(B_0)
\end{align} 
\newcommand*{\Perm}[2]{{}^{#1}\!P_{#2}}%
Assuming $\type = \type'$ and that every possible non-nullary operator application to nullary operators is defined and returns a fresh literal, $G$ grows quickly:
\begin{align}
|G_k| = \sum_{m=2}^{k+2} \sum_{f \in B_+} \binom{|G_{m-1}|}{\mathit{arity}(f)}  - \binom{|G_{m-2}|}{ \mathit{arity}(f)} \label{eq:g:grows}
\end{align}
Despite this prodigious growth rate, \Ourproject works well in practice given a
small $B$, as we show in \autoref{sec:B}.  One reason for
this is that, in our experiments, only 2\% of the applications of $f \in
B_+$ produced a new value.  

\boldpara{Indexifying a Set of Operators: }

Having described the construction of the garden of literals, $G$, that  becomes
indexified inputs and outputs for the indexified operators, we return to the
set of operators of interest that occur in $P$ and that we wish to indexify.
Let $F_+$ be this set and let  $F = G \cup F_+$ so that $G = F_0$, the set of
literals or nullary operators of interest. Note again that $F_+ \cap B_+$ may
be empty. \emph{The index function}: specify $\delta : F \rightarrow
\widehat{F}$ as an isomorphism that maps operators in $F$ to fresh names for
the indexed version of the operator that takes and returns indices over
$G$\footnote{We use $\delta$ to denote our indexing function, because
``$\delta\!\epsilon\!\iota\!\kappa\!\tau\!\eta\!\sigma$'' 
 means ``index'' in Greek and starts with $\delta$; it replaces the 
 self-explanatory, but longer and less elegant name $\mathit{indexOf}$ we
 used in the introduction.}. Generally, we
cannot index all the operators in a program, because some variables or
operators can be defined externally. 

 \begin{figure*}[t]
\centering
\begin{align} 
&\tau \in \type &\Rightarrow 
& \; ``\underline{\tau\ x}" &\rightarrow & \; ``\underline{int\ \hat{x}}'' \label{eq:type:mark} \\
& l \in G  &\Rightarrow 
& \; ``\underline{l}" &\rightarrow & \; `` \underline{\mathit{\delta}(l)}" \label{eq:lit} \\
& l \notin G\wedge \text{ typeof}(l) \in \mathcal{T} &\Rightarrow 
& \; ``\underline{l}" &\rightarrow & \; `` \underline{\bot}" \label{eq:litbot} \\
&f \in F_+ &\Rightarrow
& \; ``\underline{f}(a_1,\cdots\!,a_i,\cdots)" &\rightarrow & \; ``\underline{\hat{f}}(a_1,\cdots\!,a_i,\cdots)" %
	\label{eq:iot:call} \\
&\hat{f} \in \hat{F}_+ \wedge \exists a_i 
    \text{ s.t. } \neg \delta^?(a_i) &\Rightarrow
&\; ``\hat{f}(a_1,\cdots\!,\underline{a_i},\cdots)" &\rightarrow & %
 \;   ``\hat{f}(a_1,\cdots\!,\underline{\mathit{\delta}_\bot(a_i)},\cdots)" \label{eq:idx:arg} \\
& f \notin F_+ \wedge \exists a_i \text{ s.t. } \delta^?(a_i)
  &\Rightarrow
& \; ``f(a_1,\cdots\!,\underline{a_i},\cdots)" &\rightarrow & %
\;	``f(a_1,\cdots\!,\underline{\mathit{\delta}_\bot^{\uminus 1}(a_i)},\cdots)" %
	\label{eq:unidx:arg} \\
& f \notin F_+ \wedge \delta^?(x) \wedge \nexists a_i \text{ s.t. }  \delta^?(a_i)  &\Rightarrow
& \;``\hat{x} := \underline{f(a_1,\cdots\!,a_n)}" &\rightarrow & \; ``\hat{x} := \underline{\mathit{\delta}_\bot(f(a_1,\cdots\!,a_n))}" %
	\label{eq:idx:return}\\
& \hat{f} \in \hat{F}_+ \wedge \neg \delta^?(x) \wedge \nexists a_i \text{ s.t. }  \neg \delta^?(a_i)  &\Rightarrow
& \;``x := \underline{\hat{f}(a_1,\cdots\!,a_n)}" &\rightarrow & \; ``x := \underline{\mathit{\delta}^{\uminus 1}_\bot(\hat{f}(a_1,\cdots\!,a_n))}" %
	\label{eq:unidx:return}
\end{align}
\caption{$\Phi_S$, \Ourproject's rewriting schema: $x \in F$; $\hat{x} \in \hat{F}$; $\hat{f} \in \hat{F}_+$. We underline the redexes.}
\label{fig:rewriter:schema}
\end{figure*}

\autoref{fig:rewriter:schema} shows the rewriting schema that \Ourproject
implements to transform an input program.  For $x \in F$, let
\begin{equation}
\label{eq:idxbot}
\mathit{\delta}_{\bot}(x) = 
\begin{cases}
i    & \text{if } \mathit{\delta}_0(x) = i \\ 
\bot & \text{otherwise} \nonumber
\end{cases}
\end{equation}
In \autoref{fig:rewriter:schema}, each $a_i$ is an argument expression
and the $\delta^?$ function checks if its argument has been
directly indexed via \autoref{eq:type:mark} or converted to an index because
\autoref{eq:idx:arg} has wrapped it in a call to $\mathit{\delta}$;  when $e$ is
an expression and $\hat{x}\in\hat{F}$, its definition is:
\begin{equation} \delta^?(t) \begin{cases} \textbf{T} & \text{if } t =
``\hat{x}" \vee t = ``\mathit{\delta}_\bot(e)" \\ \textbf{F} & \text{otherwise}
\end{cases} \end{equation}

\autoref{eq:type:mark} indexes variables and function declarations, in the
latter case through repeated applications on a function's arguments and its
return type name pair. $\phi_S$ only changes the parts in the program that
appear in redexes in \autoref{fig:rewriter:schema}. For the rest of them, the
identity is implicit.   \autoref{eq:lit} replaces literals in $G$ with their
index under the homomorphism $\phi$. \autoref{eq:litbot} handles the error
case, where a constant has an indexified type but is not in $G$ by replacing it
with $\bot$.  For each function call on a function to index, \ie $\forall f \in
F_+$, \autoref{eq:iot:call} replaces the call's function identifier with the
name of its indexed variant.

\autoref{eq:idx:arg} to \autoref{eq:unidx:return} rewrite function calls, which
lack type annotations.  For this reason, they do not overlap with
\autoref{eq:type:mark}.  They handle flows between indexed regions of the
programs and unindexed ones.  \autoref{eq:idx:arg} wraps unindexed arguments to
an indexed call in $\mathit{\delta}$ to convert them to indices.
\autoref{eq:unidx:arg} it is the complement of \autoref{eq:idx:arg}: it
unindexes indexed variables that flow into unindexed functions. There are four
different combinations of the two conditions in  \autoref{eq:idx:arg} and
\autoref{eq:unidx:arg}. The two combinations that we treat are flows across
indexed to unindexed boundaries.  The other two combinations do not require
transformation as no flows across indexed and unindexed boundaries occur in
them.  \autoref{eq:idx:return} indexes returns from unindexed  ($f \notin F_+$)
functions into indexed variables ($\delta^?(x)$). The conditions $\nexists a_i
\text{ s.t. }  \delta^?(a_i) $ and $\nexists a_i \text{ s.t. }  \neg
\delta^?(a_i) $ in these two rules sequence their application, so the last two
rules only trigger after exhausting the the first six rules, as $\delta^?(a_i)$
checks if $a_i$ has been already indexed. Without the ordering, the rules
overlap and we would not be able to prove the confluence of \Ourproject's TRS
in \autoref{thrm:confl}. Similar to the previous two rules,
\autoref{eq:unidx:return} complements \autoref{eq:idx:return}: it unindexes
returns from indexed functions ($\hat{f} \in F_+$) that flow into an unindexed
variable ($\neg \delta^?(x)$). As in the previous case, we only need rules for
two out of four condition combinations (of the first two conjuncts of the
guard), as only in these two we have flows across indexed and unindexed
regions.

The schema can interact.  Consider
\autoref{eq:unidx:arg} and \autoref{eq:idx:return}. Assume we have a call to $f
\notin F_+$ that returns into a indexed variable and takes two indexed
arguments.  Two applications of \autoref{eq:unidx:arg} and one of
\autoref{eq:idx:return}, in any order, would rewrite this call.
Correctness requires \Ourproject's rewriting to be confluent.
\begin{thrm}[Confluence]
\label{thrm:confl}
All instantiations of the term rewriting schema $\Phi_S$ into term rewriting systems are confluent.
\end{thrm}
\begin{sketch}
Of the rules in \autoref{fig:rewriter:schema}, \autoref{eq:type:mark} and
  \autoref{eq:lit} share $G$; \autoref{eq:iot:call},  \autoref{eq:unidx:arg},
  and \autoref{eq:idx:return} share $F_+$; and \autoref{eq:idx:arg} and
  \autoref{eq:unidx:return} share $\hat{F}_+$. The other rules cannot overlap
  because their guards restrict them to distinct sets. \autoref{eq:type:mark}
  and \autoref{eq:lit} do not overlap because their guards partition $G$.
  Similarly \autoref{eq:iot:call} is defined over a different part of $F_+$
  than \autoref{eq:idx:arg} and \autoref{eq:unidx:arg}. The guards of
  \autoref{eq:idx:arg} and \autoref{eq:unidx:return} guarantee that they do not
  overlap: $\exists a_i  \text{ s.t. } \neg \delta^?(a_i)$ for
  \autoref{eq:idx:arg} and $\nexists a_i \text{ s.t. }  \neg \delta^?(a_i) $
  for \autoref{eq:unidx:arg}.  Finally, $\delta^?$ prevents
  \autoref{eq:unidx:arg} from being applied to eligible calls until
  \autoref{eq:idx:arg} has rewritten every  parameter within it. Thus, none of
  the rewriting schemas have overlapping terms on the left hand side and
  $\Phi_S$ is non-overlapping. $\Phi_S$ is also left-linear: no
  variable occurs more than once in the left hand sides of the rules in
  \autoref{fig:rewriter:schema}. 
  We use substitute to instantiate the term rewriting schema $\Phi_S$ into the term rewriting $\Phi_i$ for a particular program:  
  $\forall r_S \in \Phi_S, \forall t \in \mathcal{T}: \Phi_i = \Phi_S \cup r_S[t/\tau]$.
For example, let $\type = \{ \text{float}, \text{string} \}$. Then we generate the following rewriting rules from the rewriting schema in \autoref{eq:type:mark}, like 
$``\text{float } x" \rightarrow  ``int\ \hat{x}" \land ``\text{string } x" \rightarrow  ``int\ \hat{x}" $. 
Since $\Phi_S$ is left-linear and non-overlapping and the substitution does not violate either property, so is $\Phi_i$.
  Rosen~\citep{rosen1973tree} proved that
  left-linear and non-overlapping systems are confluent and so, each $\Phi_i$ is confluent, so $\Phi_S$ is confluent.\looseness=-1
\end{sketch}

\section{Implementation}
\label{sec:imp}

Given a set of operators (or functions) that can form undecidable expressions,
\Ourtool  memoizes a finite part of their behaviour, maps the rest to unknown $\bot$, then transforms a program to use them.  The
resulting program produces decidable constraints using these indexed operators
for a subset of its original state space.  We implemented \Ourproject on top of
LLVM~\citep{llvm} for the \lstinline+C+ language to produce LLVM IR for the symbolic
execution (symex) engine Klee~\citep{cadar2008klee} to execute. We use Klee's
default solver, \emph{STP}~\citep{ganesh2007decision}, because Dong \etal showed
Klee performs best with STP~\citep{dong2015studying}.  

\begin{figure*}[t]
\centering
	\includegraphics[width=\textwidth]{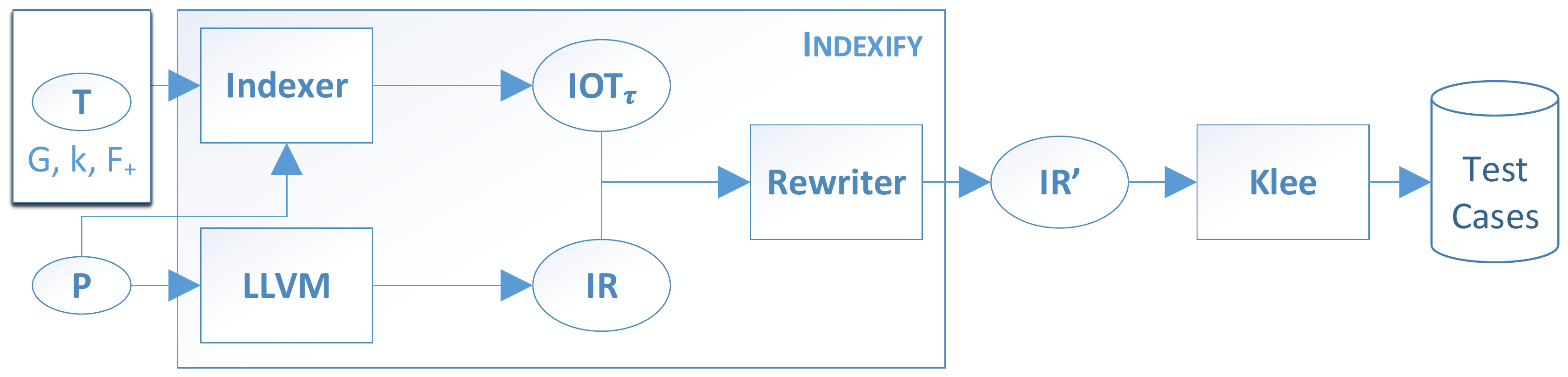}	
	\caption{The architecture of {\sc Indexify}; $G$, $k$, and $F_+$  are optional inputs.}
	\label{fig:implarchi}
\end{figure*}	

\autoref{fig:implarchi} shows the architecture of \Ourproject.  The two main
components of \Ourtool are its \textbf{Indexer} and \textbf{Rewriter}.  To use
\Ourtool, the user specifies $P$, the program to indexify, and $F_+$, the
signature of the functions (or operators) to index.  It is the user's
responsibility to ensure to specify $F_+$ so that it contains all the operators
needed to guarantee that $\textit{indexify}(P)$ produces decidable constraints,
at least along the paths the user wishes to explore.  Directly specifying $F_+$ is tedious. \autoref{sec:usage} details how we enable the user to indirectly specify $F_+$ in a simpler way.

If the user does not provide $S$, the indexer populates $S$
with constants in $P$.  Then, it computes $G$, the set of values to index and
builds indexed operator tables.  The rewriter rewrites $P$'s IR to use indexed
operators and values, including constants, and to convert indices to values and
vice versa.  Finally, \Ourtool runs Klee on the indexed IR to produce tests.   Next, we discuss the operation of the indexer and the rewriter in
detail, then explain how \Ourtool produces constraints and close with its
usage.\looseness=-1

\subsection{Indexer}
\label{sec:indexer}

For each $\tau \in \type$, we must index a subset of $\tau$'s values
($G_\tau$), but which subset?  \Ourtool is \emph{useless} without a good
selection of values from $\tau$.  For instance, if all the values in $G_\tau$
take the same path (\eg immediately exit), we learn nothing about the program
under analysis.  To build $G_\tau$, the indexer uses functions from a set of
constructors ($B_+$) to extend an initial set of seeds ($S_\tau$); obviously,
the quality of $B_+$ and $S = \cup_{\tau\in\type} S_\tau$ determine the quality
of $G = \cup_{\tau\in\type} G_\tau$.  In \autoref{sec:B}, we show
that constants harvested from the source of the input program $P$ make a 
surprisingly effective $S$ and, in \autoref{sec:eval:setup}, that
users unfamiliar with $P$ do not improve on these constants much.
Supplementing $S$ with constants from developer artefacts other than the
program text, like test cases or documentation, is future work.  If the user
does not specify $B_+$, the indexer defaults to using $F_+$, \ie $B_+ = F_+$.
\autoref{sec:good:builder} evaluates this choice of default for $B_+$.  Adding new operators in $B_+$, to evaluate using \Ourtool is straightforward: starting with the initial definitions, we automatically compute their indexed versions (Indexed Operator Table), by executing that operators, and memoising the results. We just require access to call the initial definition while memoising.

\renewcommand{\algorithmicrequire}{\textbf{Input:}}
\renewcommand{\algorithmicensure}{\textbf{Output:}}

\renewcommand\algorithmiccomment[1]{%
  \hfill\#\ \eqparbox{COMMENT}{#1}%
}%

\begin{algorithm}[t]
\caption{$\mathit{driver}(\mathit{\delta}_0,B_+,F_+,k) = \mathit{\delta}_0', \hat{F}_+$\\
This algorithm first calls \autoref{alg:extend} to build the garden $G$, then
calls \autoref{alg:memoise}, using $G$ (as carried within $\mathit{\delta}_0$) to
build the indexed operator tables.} 
\label{alg:mathit{idx}}
\begin{algorithmic}[1]
\REQUIRE \hspace*{1.3mm} $\mathit{\delta}_0$, an indexed set of nullary operators \\
\hspace*{6.35mm} $B_+$, a set of non-nullary functions for building $G$ \\
\hspace*{6.35mm} $F_+$, a non-nullary set of functions to index \\
\hspace*{6.35mm} $k$ limits applications of $b \in B_+$ \\

\ENSURE $\mathit{\delta}_0'$, an extension of $\mathit{\delta}_0$ \\
\hspace*{6.35mm} $\hat{F}_+$, the image of $F_+$ under the homomorphism $\phi$ \\[4mm]

\renewcommand{\algorithmicforall}{$\forall$\hspace*{-1mm}}
\FORALL[Build the garden $G$]{$i \in [1..k]$}
    \FORALL{$b \in B_+$}
        \STATE $\mathit{\delta}_0 := \mathit{extend}(\mathit{\delta}_0, b)$
    \ENDFOR
\ENDFOR
\FORALL[Build $\hat{F}_+$]{$f \in F_+$}
    \STATE $\hat{F}_+ := \hat{F}_+ \uplus \mathit{memoise}(\mathit{\delta}_0, f)$
\ENDFOR
\RETURN $\mathit{\delta}_0, \hat{F}_+$
\end{algorithmic}
\end{algorithm}

\begin{algorithm}[t]
\caption{$\mathit{extend}(\mathit{\delta}_0, f) = \mathit{\delta}_0'$} 
\label{alg:extend}
\begin{algorithmic}[1]
\REQUIRE \hspace*{1.3mm} $\mathit{\delta}_0$, an indexed set of nullary operators (constants) \\
         \hspace*{6.35mm} $f : \tau_0 \rightarrow \tau_1$, a function to evaluate to extend $\mathit{\delta}_0$ 

\ENSURE  $\mathit{\delta}_0'$, $\mathit{\delta}_0$ extended using $f$ \\[4mm]

\STATE \textbf{let} $m := \mathit{arity}(f)$ \textbf{in}
\STATE $G := \{ g \mid (g,n) \in \mathit{\delta}_0 \}$ 
	\COMMENT{Snapshot $G$ before changing it}
\FORALL{$g^m \in G^m \wedge \mathit{typeof}(g^m) = \tau_0$} 
        \STATE \textbf{let} $v := f(g^m)$ \textbf{in}
        \IF[\autoref{eq:idxbot} defines $\mathit{\delta}_{\bot(v)}$]{$\mathit{\delta}_{\bot}(v) = \bot$} 
            \STATE $\mathit{\delta}_0 := \mathit{\delta}_0 \cup \{(v, |\mathit{\delta}_0| + 1)\}$
        \ENDIF
\ENDFOR	

\RETURN $\mathit{\delta}_0$
\renewcommand\algorithmiccomment[1]{%
} 
\end{algorithmic}
\end{algorithm}

\begin{algorithm}[t]
\caption{$\mathit{memoise}(\mathit{\delta}_0, f) = \hat{f}$ \\ 
This algorithm builds the image of $f$ under $\mathit{\delta}_0$.} 
\label{alg:memoise}
\begin{algorithmic}[1]
\REQUIRE \hspace*{1.3mm} $\mathit{\delta}_0$, an indexed set of nullary operators \\
\hspace*{6.35mm} $f : \tau_o \rightarrow \tau_1$, a non-nullary function to index \\

\ENSURE $\hat{f}$, the image of $f$ under the homomorphism $\phi$ \\[4mm]

\STATE \textbf{let} $m := \mathit{arity}(f)$ \textbf{in}
\STATE $G = \{ g \mid (g,n) \in \mathit{\delta}_0 \}$
\STATE \textbf{let} $\hat{f} = \{\}$ \textbf{in}
	\COMMENT{$\hat{f}$ is the image of $f$}
\FORALL{$g^m \in G^m \wedge \mathit{typeof}(g^m) = \tau_0$}
        \STATE \textbf{let} $v := f(g^m)$ \textbf{in}
        \STATE $\hat{f} := \hat{f} \cup \{ ((\textbf{map}\ \mathit{\delta}_0\ g^m),\mathit{\delta}_0(v))\}$ 
\ENDFOR

\RETURN $\hat{f}$
\end{algorithmic}
\end{algorithm}

Starting from $S$, \Ourtool first constructs $G$ using the functions in $B_+$,
as specified in \autoref{alg:mathit{idx}}.  It loops over number of applications $k$
and over $B_+$, calling \autoref{alg:extend}.  Later calls to
\autoref{alg:extend} are applied to the results of early calls.
In this section, we split $\delta$ (\autoref{sec:formalism}) into $\delta_0$ for the nullary operators and $\hat{F}_+$ for the rest of the operators. \autoref{alg:extend} builds $\mathit{\delta}_0 = \phi|_{F_0}$ and \autoref{alg:memoise} builds $\hat{F}_+$.\looseness=-1

\autoref{alg:extend} simply enumerates $G^m$, filtering out type invalid
permutations.  In future work, we plan to experiment with sampling $G^m$.
Given $S = \{a,b\}$, $K=2$, and $B_+ = \{ \mathit{strcat}, \mathit{strstr} \}$, \Ourproject
produces $G = \{ \emptyset, a, b, aa, bb, ab, ba \}$. 
Here, $G$ contains all the values that can be obtained when applying the
functions $\mathit{strcat}$ and $\mathit{strstr}$ twice first on $S$, then on $S$ and
the result of the first application.  As discussed in \autoref{sec:formalism},
$G$ grows quickly.  Despite this forbidding growth rate, our experiments in
\autoref{sec:B} show that \Ourtool performs effectively
when restricted to small values of $k$ and $|B|$.\looseness=-1   

Next, \autoref{alg:memoise} indexes $F_+$, the functions operating over $\tau
\in \type$.  If the user inputs \type,
we harvest $F_+$ from the intersection of functions used in the program, and a
relevant library, \ie \lstinline+string.h+, when indexing strings, and
\lstinline+maths.h+, when indexing floats or double.  For each $f : \vec{\tau}
\rightarrow \tau \in F_+$, the indexer encodes $\hat{f}$ (or \lstinline+i_f+ in
ASCII), the indexed version of $f$, as a sequence of if statements as follows:\looseness=-1
\begin{lstlisting}[style=codeSnippet] 
i_f(@$\tau$@ x, @$\tau$@ y) { 
   if x = 1 @$\land$@ y = 1 return 1; 
   if x = 1 @$\land$@ y = 2 return 3; 
   ...
   return -1; // return @$\bot$@ when x @$\notin G$@ @$\lor$@ y @$\notin G$@
}
\end{lstlisting}
Indexed operators (or functions) memoize the result of $f \in F_+$ on values in
terms of the indices defined by $\mathit{\delta}$.  When the result of $f$'s
computation is not in $G$, $\hat{f}$ returns $\bot$.  Thus,
\lstinline+i_f+ above contains \lstinline[style=codeSnippet]+if x = 1+ $\land$
\lstinline+y = 3 return 57;+ 
because $f(\mathit{\delta}^{\uminus 1}(1),
\mathit{\delta}^{\uminus 1}(3)) = \mathit{\delta}^{\uminus 1}(57)$.  The number of if statements in
an indexed operator is $|G|^n+1$, but
Sections~\ref{sec:bound:k}~and~\ref{sec:B} show experimentally that
\Ourproject was effective using parameter settings that only doubled the size
of its inputs on average.  The indexer injects the definition of each $\hat{f}
\in \hat{F}_I$ into $P$.\looseness=-1

Algorithms 1--3 build $\delta$ and exactly realise \autoref{eq:buildG}.
\autoref{alg:mathit{idx}} is the driver algorithm that calls
\autoref{alg:mathit{idx}} and \autoref{alg:extend} to build $\delta = \delta_0
\cup \delta_+$ in two parts. \autoref{alg:extend} simultanously extends
$\delta_0$ and constructs the garden $G$ using the builders $B_+$ applied to
$G$ as it expands. \autoref{alg:mathit{idx}} calls \autoref{alg:extend} only
over $B_+$ on lines 1--3.  \autoref{alg:extend} calls each $b \in B_+$ on the
values currently in $G$. If the resulting string is not currently in $G$,
\autoref{alg:extend} assigns it a fresh index and updates $\delta_0$.
\autoref{alg:memoise} constructs $\delta_+ = \hat{F} - \hat{F}_0$, the union of
all the indexed operator tables for the functions in $F_+$.
\autoref{alg:mathit{idx}} calls \autoref{alg:memoise} only over $F_+$ on lines
4--5. \autoref{alg:extend} and \autoref{alg:memoise} perform no other
computations.

\subsection{Rewriter}
\label{sec:rewriter}

The \emph{rewriter} is a straightforward implementation of the term rewriting schema in \autoref{fig:rewriter:schema}. We generate wrapper code that unindexes things on the fly. We initially index every occurrence of a variable of the indexed type, and further we unindex them on the fly, when the variables flow into unindexed operations. In the presence of aliasing, we unindex such values when they go into a different data type (structure), and later unindex it back, when it flows into an indexed data type / operator. To handle casts we use the index, and unindex function. When the program casts a indexed variable $x$ to a different type, we first unindex $x$ to the initial data type, and then we cast the variable of the initial data type, as in the initial program. Similarly, when the program casts a variable of an unindexed type to an indexed type, we index the result of the cast applied on the initial (unindexed) data type.\looseness=-1

 \autoref{fig:indexification:ex} shows an example of applying $\phi_S$. When $G=\{$ "\lstinline{foobar}", "\lstinline{oobar}", "\lstinline{bar}", "\lstinline{oo}" $\}$, \Ourtool
replaces "\lstinline+bar+" on line 6  (\autoref{lst:P}) with $\mathit{\delta}($"\lstinline+bar+"$) = 2$ (\autoref{lst:P'}), but ignores the constants \lstinline+a+ and \lstinline+foo+
on line 5 (\autoref{lst:P}). \autoref{eq:iot:call}  replaces \lstinline+strstr+ on line 6 in \autoref{lst:P} with
\lstinline+i_strstr+ on line 14 in \autoref{lst:P'}.  \autoref{eq:unidx:arg} unindexes indexed variables that flow into
unindexed functions.  In \autoref{lst:P'}, \autoref{eq:unidx:arg}  unindexes
the indexed variable \lstinline+S1+ on line 13, as the function
\lstinline+puts+ is unindexed. Finally, \autoref{eq:idx:return} indexes the return from unindexed functions into indexed variables  in
\autoref{lst:P'} on line 11 since \lstinline+strcat+ is not indexed, but returns
into \lstinline+S2+, which is. 

The rewriter uses
\lstinline+LLVM+'s API for IR manipulations and transformations\footnote{\url{http://llvm.org}}. In
\lstinline+LLVM+, types are immutable, so we cannot change them in place. Instead,
\Ourproject outputs the indexed version of the IR of $P$ in a new file.  The \emph{rewriter} is a  visitor that walks the original IR
of $P$: when it reaches an IR element of a type
in \type, it creates a new instruction with \type changed to
\lstinline+index+; otherwise, it simply echoes the instruction.  In LLVM, a global variable and its
initializor must have the same type and LLVM forbides casts or
function calls in initializors.  Thus, the visitor replaces values with indices in
initializors.  We execute Klee on this indexed version of the LLVM bitcode.  To
support indexing multiple types at time, \Ourtool keeps multiple gardens, one
for each indexed data type.  For example, one might want to index both the strings and the floats in a program. For this, \Ourtool keeps $G_S$, the set of string values in our domain restriction, and $G_F$, the set of float values in our domain restriction. The indexes in $G_S$ and $G_F$ do not overlap. Further, \Ourtool proceeds normally, but when indexing a type it uses indexes from the corresponding garden: $G_S$ if the type to index is a string; $G_F$ if the type to index is a float.

\Ourproject allows an indexed operator to take
both indexed and unindexed types. In this case, \Ourproject does not simply
replace the function's body with an $\mathit{IOT}$;  instead, it indexifies the
function's body, replacing any internal calls to indexed operators with their
indices versions and inserting index-casts to (un)index values, including the
return value, as needed. Currently \Ourtool does not use \autoref{eq:litbot}. \Ourtool does not need \autoref{eq:litbot} because for us $S$ is the set of the literals in the program. Thus, all the literals in the program are in $G$.

\subsection{Indexed Constraint Construction} 
\label{sec:constraint}

\begin{figure}[t]
\begin{subfigure}[b]{0.35\textwidth}
\begin{lstlisting}[style=codeSnippet]
int main(){
	char S1[3], S2[5];
	klee_make_symbolic(S1, sizeof(S1), "S1");
	puts(strlen(S1));
	S2 = strcat("a","foo");
	if(strstr(S1, "bar")) return 1;
	else return 0;
}
\end{lstlisting}
\caption{P, the program to indexify.}
\label{lst:P}
\end{subfigure}\hfill
\begin{subfigure}[b]{0.6\textwidth}
\begin{lstlisting}[style=codeSnippet]
int  i_strstr(int s1, int s2){
	if(s1 == 0 && s2 == 2) return 2;
	if (s1 == 0 && s2 == 3) return 1;
	if (s1 == 3 && s2 == 2) return 0;
	...	return -1; // return @$\bot$@
}
int main() {
	int S1,S2;
	klee_make_symbolic(&S1, sizeof(S1), "S1");
	puts(@$\delta^-1(S1)$@);
	S2 = @$\delta$@(strcat("a","foo"));
	if(i_strstr(S1, 2)) return 1;  // 2 = @$\delta$@("bar");
	else return 0;
}
\end{lstlisting}
\caption{P' = indexify(P, string); \lstinline+puts+ and \lstinline+strcat+ are not in $F_+$.}
\label{lst:P'}
\end{subfigure}
\caption{Indexification: \Ourtool also injects $\mathit{\delta}$ and $\mathit{\delta}^{\uminus 1}$ into $P'$, although not shown.}
\label{fig:indexification:ex}
\end{figure}

To understand how an indexified program constructs constraints, consider the
program $P$ in \autoref{lst:P}\footnote{This is an actual, if pedagogical
example.  We provide the constraints in SMTLIB format, as generated by Klee, at <url>.}. Here, the string $S$ is symbolic; and ``bar''
is a constant string. The call to \lstinline+klee_make_symbolic+ makes the
program variable $S$ symbolic. When $G=\{$ ``foobar'', ``oobar'', ``bar'', ``oo'' $ \}$,
\lstinline@indexify --type string P.c --garden path_to_G --F_+ path_to_F_+@ produces $P'$ in \autoref{lst:P'}, whose behaviour is 
restricted to $G$ over string operations.  In $P'$, $\hat{f}_\text{strstr} = \mathit{i\_strstr}$.

\Ourtool produces $P'$.  When we symbolically execute $P'$, we reach the
call to  \lstinline+i_strstr+ on line 12 (\autoref{lst:P'}). For \lstinline+i_strstr+, 
Klee delays
calling the internal solver~\citep{cadar2008klee}, encoding
\lstinline+i_strstr+ into an indexed operator table ($\mathit{IOT}$) 
as disjunctions 
for the branches of the \lstinline+if+ statement (\autoref{lst:P'}). 
The constraints that \lstinline+i_strstr+ generates are:
\lstinline[mathescape]+(s1 = 0 $\land$ s1 = 2 ) $\lor$ (s1 = 0 $\land$ s2 = 3) $\lor$ (s1 = 3 $ \land s2$ = 2) $\lor$ $\mathit{IOT}$+.

The constraints in the $\mathit{IOT}$ of $\hat{f} \in \hat{F}_+$ are in
the theory of equality by construction;  an $\mathit{IOT}$ is a lookup table
that disjuncts equalities over the values of its parameters.  These constraints
are solvable by an integer solver equipped with equality theory in polynomial
time. Barrett \etal~\citep{barrett2009satisfiability} show this and provide a linear time solving algorithm.  Our implementation currently
cannot fully exploit this fact because it extends Klee, which builds constraints
directly in the bitvector theory with arrays. In our current implementation, \Ourtool first indexes the program then uses Klee for symbolic execution. Klee takes the indexed program and bitblasts it. In this case, \Ourtool still improves performance whenever bitblasting integers results in shorter constraints than bitblasting the initial data type.  For example, a string of length $N$ requires $N * \mathit{sizeof}(\text{char})$ bits. After indexing, the same string uses only $\textit{sizeof}(\text{int})$ bits.

\Ourproject takes advantage of Klee's internal query optimisations. Klee  
removes unsatisfiable constraints from the path conditions via 
simplifications~\citep{cadar2008klee, dillig2010small}. Under these optimizations, Klee selects only
the relevant subset of $\hat{f}_\text{strstr}$ in the context of the predicate
in the if statement on line 15 in \autoref{lst:P'}: only the entries where the
second parameter is $\mathit{\delta}(\text{bar}) = 2 $. It does so prior to
sending the query to the internal solver. Thus, the constraints that Klee
actually sends to STP for \lstinline+i_strstr+ are: 
\lstinline[mathescape]+(s1 = 0 $\land$ s1 = 2 ) $\lor$ (s1 = 3 $\land$ s2 = 2) $\lor$ $\mathit{IOT}|_{S_2=2}$+.

\autoref{sec:clauses} compares the number of clauses that Klee and \Ourproject
generate on coreutils. Without optimisation, \Ourproject generates three times
more clauses than Klee. The results drastically change, when we look at the
number of clauses that reach the solver, after Klee's internal optimisations:
indexified programs send $\frac{1}{10}$ as many constraints to the solver.
\Ourproject's $\mathit{IOT}$, by construction, are particularly amenable
to the syntactic constraint simplifications that Klee employs.

\subsection{Usage} 
\label{sec:usage}

Once Klee is successfully installed, \Ourproject is easy to use:  issuing
\lstinline+indexify --type string yourprogram.c+ indexifies
\lstinline+yourprogram.c+, and then runs Klee on the indexified program.
\Ourproject takes optional parameters.  When  runs with its
\lstinline+--outputIndexedIR+ flag, \Ourtool outputs the indexed IR.  By default,
\Ourproject harvests $S$, the set of seed values to index  from the input program.  The switch \lstinline+--addSeeds <file>+
specifies  a file containing seeds to add to the harvested
constants; \lstinline+--seeds <file>+ specifies a file containing
$S$ and prevents constant harvesting.  By default, \Ourproject automatically
constructs the operator definitions, using \autoref{alg:memoise}.  The switch
\lstinline+--indexOpDefs <file>+ specifies the  \lstinline+LLVM+
bitcode file that contains indexed operators to allow their reuse across runs,
amortizing the cost of their construction. 

Specifying $F_+$, the functions in a program to index, is tedious. We
allow the user to specify only \type,  which triggers the indexer to populate
$F_+$ from a header file.  For \type = \lstinline+string+,
\Ourtool indexes all the functions in \lstinline+string.h+; for \type = \lstinline+float+, \Ourtool indexes all the functions in
\lstinline+maths.h+. For an arbitrary data type, the user needs to specify the
name of the desired header file.

\section{Experimental Setup}
\label{sec:eval:setup} 

\Ourproject operates on an input program $P$ in two phases.  First, its indexer
constructs the ``garden'' of values $G$, then it builds operator tables for the
target operators in $P$ over $G$.  Finally, \Ourproject transforms $P$ to $P'$ and
symbolically executes it.  As detailed in the previous section, the indexer
takes four parameters:  the types to index or a list of functions to index
$F_+$, the seed values $B_0 = S$, the functions for building the garden from
$S$, and $k$, a bound on the applications of the builder functions.  In this
section, we explore various setting for $S$, $B_+$, and $k$ in order to 
fix them, leaving only $F_+$ to vary, in \autoref{sec:res}.\looseness=-1

\boldpara{Corpus}
Our experimental corpus contains two benchmarks: 
\lstinline+GNU Coreutils+\footnote{\url{http://www.gnu.org/software/coreutils/coreutils.html}},
and
\lstinline+fdlibm53+\footnote{\url{http://www.validlab.com/software/fdlibm53.tar.gz}}.
GNU Coreutils contains the basic file, shell and text manipulation utilities
of the GNU operating system, such as \lstinline+echo+, \lstinline+rm+,
\lstinline+cp+ and \lstinline+chmod+.  We include all the 89
\lstinline+Coreutils+ in our experiments, and use
\Ourproject$(\mathit{string})$ on them. We use the second benchmark,
\lstinline+fdlibm53+, for exploring the capabilities of \Ourproject, when
indexifying floats.  We picked these benchmarks because
\Ourproject, Klee and Zesti can execute all the benchmarks in our corpus.

We select Klee (KLEE 1.1.0 47a97ce; LLVM 2.7), as a dynamic symbolic execution engine, and Zesti (the beta version on the Zesti's website\footnote{\url{http://srg.doc.ic.ac.uk/zesti/zesti.tar.gz}}) as a concolic testing engine, for the comparison with \Ourtool. We selected these 2 tools, as \Ourtool is based on Klee, as is the case for Zesti. \Ourtool first indexifies the program, and further calls Klee on the indexified program. We also decided to compare \Ourtool with Klee and Zesti, because of the fact that we support symbolic execution on \lstinline+C+ code. Klee is the most famous symbolic execution engines for \lstinline+C+ code. We called all the three tools (\Ourtool, Klee, and Zesti) with exactly the same parameters as used in the initial Klee paper~\citep{cadar2008klee}. 

\subsection{Human-Provided Seeds} 
\label{sec:evalg}

\Ourproject constructs $G$, the subspace of value to which \Ourproject
restricts a program's behaviour, from $S$.  Thus, defining $S$ is crucial to
\Ourproject's effectiveness.  \Ourproject defaults $S$ to the constants of
\type in $P$.  This extraction method exploits the domain knowledge embedded 
in these constants to bias $B$, and therefore $G$, to values that $P$ is 
more likely to compute.

Here, we ask whether human intuition can help \Ourtool by augmenting $S$ with
values that a developer considers interesting?  To decide if human intervention
in seed selection is effective, we compare the manual effort to discover seeds
against the coverage and execution time gains of the augmented set of seeds.
For this comparison, we uniformly selected 10 programs from
\lstinline+coreutils+: \lstinline+cat+, \lstinline+expand+, \lstinline+fold+,
\lstinline+mknod+, \lstinline+mktemp+, \lstinline+runcon+, \lstinline+shred+,
\lstinline+tsort+, \lstinline+unexpand+, and \lstinline+wc+. One of the authors
spent no more than ten minutes to construct strings that he thought might allow
symbolic execution of the indexed program to explore new paths or corner cases
and added them to $S$.  We ran \Ourproject with $T$ set to strings, $B_+$ set
to string functions occuring in $P$ and uclibc, and $k = 3$ on this set.  

The results were identical:  the human-augmented seeds provided no discernable
improvement.  Given \Ourproject's overall effectiveness, we take this as a
testament to the effectiveness of \Ourproject's constant harvesting heuristic.

\subsection{Bounding $k$ to Operator Chain Length} 
\label{sec:bound:k}

\Ourproject's core indexer algorithm (\autoref{alg:mathit{idx}}) is
computationally and memory expensive as a function of $k$, the number of
applications of the functions in $B$ (\autoref{eq:g:grows}).  Setting $k$ is a
trade-off between the power of analysis (\ie the size of $G$) and the computational cost of
running \Ourtool.  First, we identify the
value of $k$ in our corpus, then we observe the performance when using the identified $k$.

From our corpus, we infer $k$ from the lengths of chains of applications over
$F_+$; we define chains have nonzero length.  In our first experiment, we
statically symbolically execute our corpus to collect symbolic state, from
which we extract operator chain lengths from def-use chains in killed
expressions.  \autoref{fig:kreal} shows the lengths of operator chains $k$ in
our corpus.  The first boxplot, labelled \emph{All.coreutils}, reports the
distribution of all operator chain lengths in \lstinline+coreutils+: the median
is 11; the minimum is 1; and the maximum is 2833. The percentage of outliers is
11.66\% upward and 0\% downward, since 1 is the first quartile and we do not
have 0s. A manual exploration of uniformly picked outliers reveals that loops
cause them. 

\begin{figure}[t]
\centering
    \includegraphics[width=0.5\columnwidth]{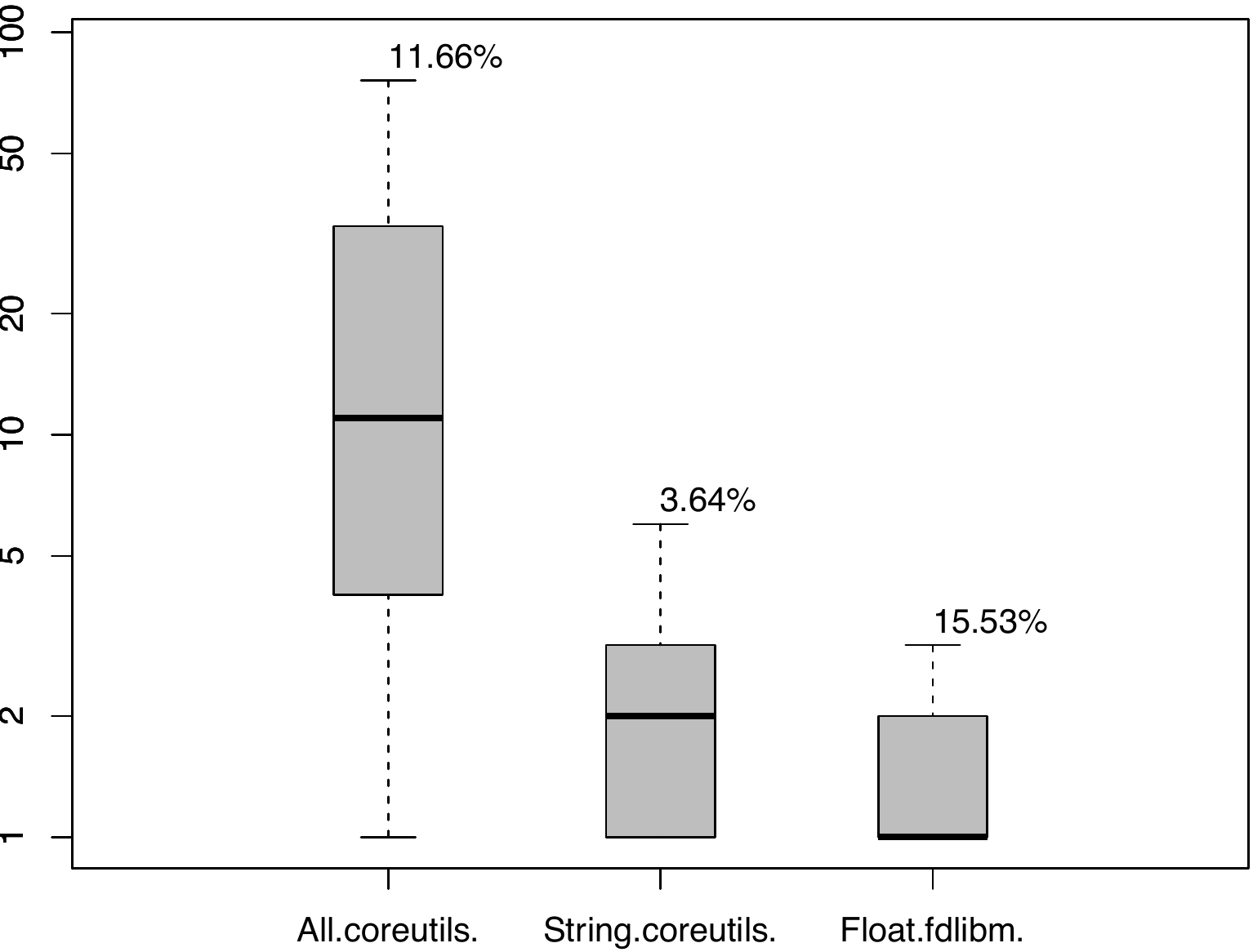}
      \caption{Length of operator chains in coreutils; we use these values to set $k$.} 
    \label{fig:kreal}%
\end{figure}

\autoref{fig:kreal} also reports boxplots for \lstinline+string+ and
\lstinline+float+ operators.  For \lstinline+string+ operators, the median
value is 2, with 3.64\% outliers; for \lstinline+float+ operators, the median
value is 1, with 15.53\%  outliers.  Setting $k$ to the third quartile of
operator chain lengths in our corpus reduces the probability that symbolically
executing an indexified benchmark will escape $G$ and bind $\bot$ to a
variable, merely through operator applications.  Thus, we fix $k = 3$.\looseness=-1

\subsection{Finding a Good Builder}
\label{sec:good:builder}

\Ourtool uses the functions in $B_+$ to build the garden $G$ from $S$. Which
functions should we use?  For strings, we consider two different builders.
$B_+^* = \{ {}^* \}$ contains only Kleene closure and $B_+^\circ = F_+$,
the indexed functions in $P$.  To build $G$, we apply the operators in $B_+$
up to $k$ times.
For example, let $S = \{a.b\}$ and $k = 2$.  Under $B_+^*$, $G = \{ \emptyset, a, b,
aa, bb, ab, ba \}$.  For floats, $B_+^\circ$ applies all the float operators in
$P$ including \lstinline+maths.h+ on all combinations of float constants in the
program text up to $k$ times.  Each application of $B_+$ potentially generates
a new value in $G$.

\begin{table}[t]
  \caption{Measures of \autoref{alg:mathit{idx}} as a function of $k$.}
\label{tab:kVariance}%
\small
\begin{tabular}{r r r r r r r}
\toprule 
& \multicolumn{2}{c}{$k = 1$} & \multicolumn{2}{c}{$k = 2$} & \multicolumn{2}{c}{$k = 3$} \\
& $B_+^*$ & $B_+^\circ$ & $B_+^*$ & $B_+^\circ$ & $B_+^*$ & $B_+^\circ$ \\
\midrule
\textbf{Time(sec)} & 324 & 123 & 544 & 263 & 1008 & \textbf{363}  \\
$|G|$  & 10792 & 69068 & 78920 & 163378 & 356760  & 263616 \\
\midrule
\textbf{ICov(\%)}  & 26.22 & 52.21 & 34.09 & 55.65 & 55.32 & \textbf{57.67}\\
\textbf{BCov(\%)}  & 21.93 & 59.74 & 39.98 & 62.37 & \textbf{69.73} & 66.83\\
\bottomrule 
\end{tabular}
\end{table}

To compare $B_+^*$  and $B_+^\circ$ on strings up to $k = 3$, we 
executed \Ourtool on
coreutils, then ran Klee on the resulting indexed programs.  We constrained the $G$ that $B_+^*$ constructed: only adding a
generated value to $G$ if its length was less than or equal to 8. We experimented with different maximum sizes, up to 20. We observed that the branch and statement coverages are not improving any more when considering strings with lengths greater than 8. Our experimental corpus, \lstinline+Coreutils+, does not computes during its execution strings longer than 8 character. This is because the most of the programs in \lstinline+Coreutils+ do not use very often the concatenation operator over strings \lstinline+strcat+. This is the only string operator that increases the sizes of strings during execution. The most of the strings that the programs in \lstinline+Coreutils+ use are flags, or file names, which are in general short in size. For different experimental programs, higher values of maximal string lengths might be required.
\autoref{tab:kVariance}'s ``Time'' row shows
that $B_+^*$ executes more than three times slower than $B_+^\circ$.
At $k=3$, 
$B_+^*$ covers 55.32\% of the instructions in coreutils (``ICov''), while
$B_+^\circ$ covers 57.67\% of them. The branch coverage (``BCov'') is
69.73\% for $B_+^*$ and 66.83\% for $B_+^\circ$.  These results show a
trade-off between execution time and coverage.  $B_+^*$ covers more
strings, increasing both \emph{BCov} and the execution time.  The difference
in \emph{BCov} is only 3\% and $B_+^\circ$ achieves 2\% higher ICov.  Thus,
we set $k=3$ and $B_+ = B_+^\circ$ in \autoref{sec:res}.\looseness=-1

When using $k$ greater than $3$, we observed an increase in runtime and a decrease in code coverage. This is due to the fact that the number of conjunctions in the constraints that \Ourtool generates grows exponentially, and Klee's internal solver time-out is reached. When this happens, Klee abandons the path with the constraint that time-outs. We ran $k$ up to $10$. The branch coverage decreased smoothly from $69.73\%$ when $k = 3$ to $65.00\%$ when $k = 10$. The execution time increased smoothly from $1008$ seconds when $k = 3$ to $3150$ when $k = 10$.\looseness=-1 

\autoref{tab:kVariance} shows that, for both
$B_+^\circ$ and $B_+^*$, coverage increases with $k$ up to $3$, but
execution time does not become unreasonable. 
Even $k = 3$ generates an immense garden, as \autoref{eq:g:grows}, in \autoref{sec:terminology} shows and $|G|$ in \autoref{tab:kVariance} confirms.
How does \Ourproject manage to be effective, despite \autoref{eq:g:grows}?
We hypothesized that 
$k > 3$ is feasible because very few elements are unique in practice.
Especially with strings, many operators
return substrings of existing elements. 
We evaluated our hypothesis on $B_+^\circ$.
Our results show that from the total of generated values, only 1.97\% are
unique; only the unique values are indexed.

\section{Evaluation }
\label{sec:res}

Here, we demonstrate that \Ourproject can improve existing automated testing techniques by trading space to reduce time and increase solution coverage. 
When used to generate test data, dynamic symbolic execution and concolic testing aim to achieve the highest structural coverage possible. 
Since the {\em theoretical} space complexity of \Ourproject is worse than exponential, an essential goal of this evaluation is to demonstrate that its space consumption, in practice, is manageable.
Furthermore, we evaluate the degree to which this manageable increase in space consumption reduces time and improves solution coverage. We evaluate whether \Ourproject can catch bugs that are out of reach to traditional symex. Then, we assess \Ourtool ability to make the constraints easier for the underlying SMT solver by restricting the domain of some operators. Finally, we evaluate \Ourproject when indexing floats.\looseness=-1

\subsection{Trading Space for Time and Coverage}
\label{sec:B}

\begin{wraptable}{l}{7cm}
\caption{Overall results for \Ourproject(String).  We fixed the bound $k=3$;  we used $B_+ = B_+^\circ$. The trends support our core insight: \Ourproject increases memory consumption reducing execution time and increasing code coverage }
\label{tab:results} 
\small
\centering
\begin{tabular}{r r r r}
\toprule 
\multicolumn{1}{c}{\textbf{Metric}} & \multicolumn{1}{c}{\textbf{\Ourtool}} & \multicolumn{1}{c}{\textbf{Klee}} & \multicolumn{1}{c}{\textbf{Zesti}} \\
\midrule 
Time(min) & 6.05  & 49.52 & 22.70\\
Memory(MB) & 900.45 & 241.16 & 3000.00 \\
ICov(\%) & 57.67 & 41.24 & 20.60\\
BCov(\%) & 66.83  & 30.10 & 14.79\\
\bottomrule 
\end{tabular}
\end{wraptable}

\Ourtool trades memory for reduced execution time and increased code coverage.
Memory has become cheaper and more abundant, but our core algorithm consumes
exponential memory in the worst case. Can we significantly reduce execution
time or increase code coverage at reasonable memory cost?

To answer this question, we compare \Ourproject to dynamic symbolic execution
using Klee and to concolic testing using Zesti.  Zesti uses concrete inputs to kick-off concolic testing. It searches a neighbourhood around the path executed under the concrete inputs. For each concrete input, Zesti follows the concrete execution that the input generates and records the path condition. When reaching  exit, Zesti  backtracks to the nearest condition, complements it, generates a new input that obeys the  new path condition, and  restarts execution from the entry point~\citep{godefroid:dart, sen2005cute}.

We compare their performance
in terms of run time, memory consumption, instruction, and branch
coverages.  We index strings in the input programs when running \Ourtool, \ie
$\type = \text{string}$.

To configure \Ourproject, Klee, and Zesti, we use the same
settings that Klee used on the coreutils benchmark~\citep{cadar2008klee}. The
time-out is 3600 seconds; the maximum memory usage allowed is 1000 MB; and the
maximum time spent on one query is 30 seconds. Although the maximum memory
consumption is 1000 MB, Zesti might use more memory than this, as it
composes multiple symbolic execution runs.  We use $k=3$, as we explain in
\autoref{sec:bound:k}.

We construct $IOT$ and $G$ online. We automatically harvest the seeds $S$
from each program's text. $F_+$ contains the functions involving strings in the
input program; we use $B_+^{\circ,3}$ (\autoref{sec:eval:setup}) to build $G$.\looseness=-1

We report the instruction and branch coverage on \lstinline+LLVM+ bitcode, for
the final linked bitcode file. This bitcode combines tool and library code.
Achieving 100\% coverage of this bitcode is usually impossible, because
programs tend to use very little library code, relegating the rest to dead
code.  For example, \lstinline+printf("Hello!")+ does not cover
\lstinline+printf+'s format specification handling code~\citep{cadar2008klee}.

\autoref{tab:results} reports our experimental results. As expected, these
results (row ``Memory'') show that \Ourproject consumes more memory than Klee.
Averaged over all coreutils, Klee requires 241.16 MB, while \Ourtool requires
900.45 MB. \Ourproject's memory consumption is reasonable relative to Zesti,
consuming $\frac13$ the memory (3000MB) Zesti does.  Zesti exceeds the 1000MB
memory limit on a single run of Klee,  as it runs Klee multiple
times.\looseness=-1

\Ourproject has compensatory advantages to set against its memory consumption.
First, \Ourtool improves time performance.  Over all, \Ourtool requires 6.05
minutes mean time, while  Zesti requires  22.70 minutes and Klee requires 49.5
minutes.
Second, \Ourproject outperforms the existing state of the art in terms of coverage achieved: \Ourtool achieves 57.67\% instruction coverage and 66.83\% branch coverage. By contrast, Klee achieves 41.24\% instruction coverage and  30.10\% branch coverage while 
Zesti achieves 20.60\% instruction coverage and 14.79\% branch coverage. 

While there is much debate in the testing community over the value 
coverage~\citep{gligoric:comparing,kletal:taicpart09},
there is little doubt that greater coverage is to be strongly preferred over lower coverage. We argue that the increased coverage and reduced run-time are more significant than the increase in memory consumption. For an increase of 659.29 MB of memory for \Ourtool,  we get 16\% increase in instruction coverage; a 36\% increase in branch coverage; and a 43 minute reduction in run time, on average.\looseness=-1

\subsection{Unique Statements Coverage}
\label{subsec:unique}

\begin{figure}[t]
\centering
\includegraphics[width=0.65\columnwidth]{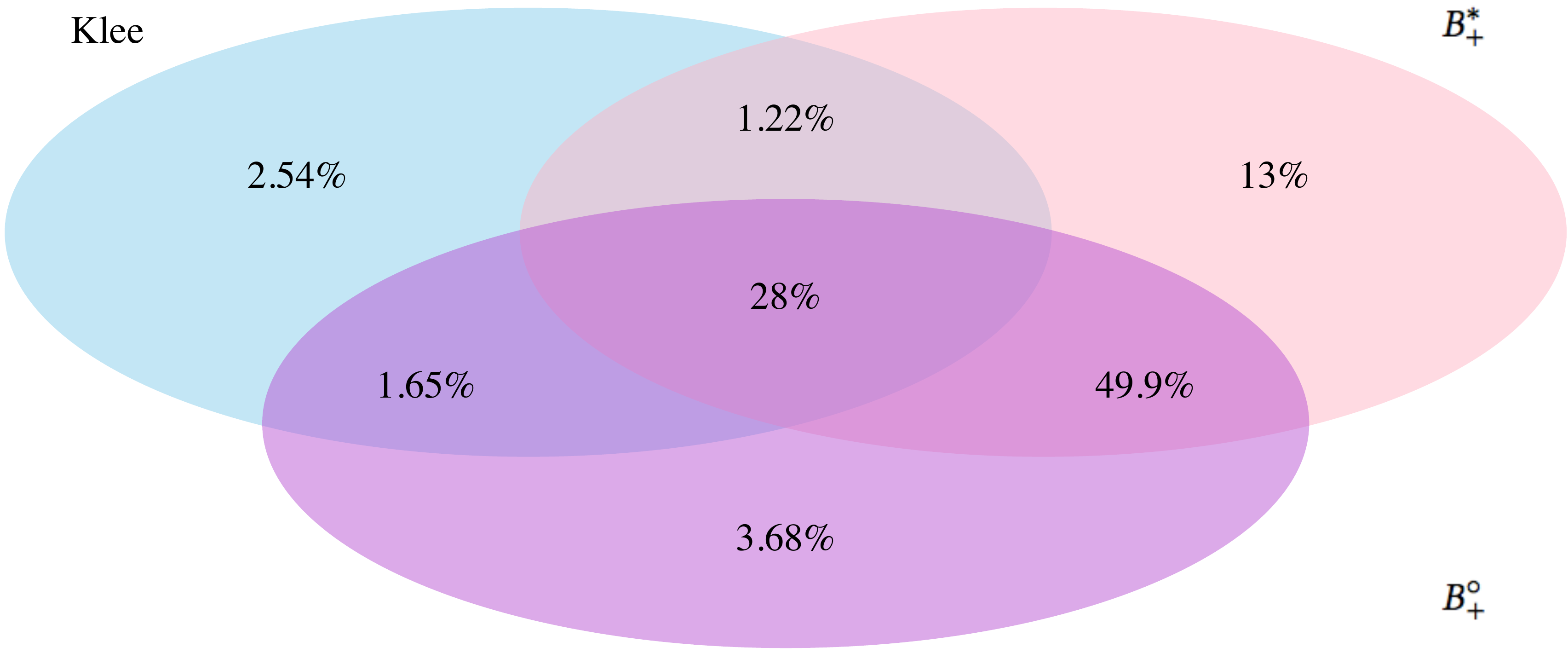}

	\caption{Uniquely covered branches by: Klee, \Ourproject with $B_+^*$, and \Ourproject with  $B_+^\circ$} 
	\label{fig:venn}
\end{figure}	

We explore how many previously unreachable branches \Ourproject brings in the reach of symex and compare how many branches Klee and \Ourtool uniquely cover. In \autoref{fig:venn}, we show the percentages of branches that  $B_+^\circ$,  $B_+^*$, and Klee uniquely cover. Klee uniquely covered 3.02\% of branches in \lstinline+coreutils+'s IR. Both $B_+^\circ$ and  $B_+^*$ missed these branches: they are infeasible in the indexed programs, because our garden $G$ does not contain the strings that these branches use. To cover these branches, a greater $K$ might help.\looseness=-1 

\autoref{fig:venn} shows that although 2.54\% of statements are covered only by \lstinline+Klee+, \Ourproject enables symbolic execution to cover far more previously out of reach branches: on total, 66.58\% are only covered by $B_+^*$, or $B_+^\circ$. $B_+^*$ covers uniquely the most branches: 13\%. $B_+^\circ$ covers uniquely 3.68\%. More branches are uniquely covered by the $B_+^*$ because the \emph{Kleene} closure contains strings that are not the result of operator applications. Some branches are uniquely covered by $B_+^\circ$ because $K$ applications of string operators may lead to strings longer than the bound for the \emph{Kleene} closure.\looseness=-1

\subsection{Bug Finding}

\begin{table}[t]
\caption{Bugs that Klee, and \Ourtool can reach.}
\label{tab:unreachableBugs} 
\small
\centering
\begin{tabular}{c r r c c c}
\toprule 
\textbf{Project} & \multicolumn{1}{c}{\textbf{Bug}} & \multicolumn{1}{c}{\textbf{Size}} & \textbf{Klee} & \textbf{\Ourproject} \\
\midrule 
ncompress & stack smash & 1935 &  NO & YES \\ 
gzip & buffer overflow  & 4653 & NO & NO \\ 
man & buffer overflow & 2805 & NO & YES \\ 
polymorph & buffer overflow & 240 & YES & YES \\ 
\bottomrule 
\end{tabular}
\vspace*{-3mm}
\end{table}

The most compelling evaluation for any testing technique is the ability to
reveal bugs that other techniques cannot. Accordingly, we investigate whether
\Ourproject can identify bugs not detected by Klee.\looseness=-1 We evaluated
\Ourproject on a set of bugs from \emph{Bugbench}, a famous benchmark for C bug
finding tools~\citep{lu2005bugbench}.  We consider only the  bugs that the
default oracle in Klee can detect --- bugs that cause programs to crash.  For
Klee, experiments~\citep{cadar2008klee} on \lstinline+coreutils+ set the
time-out to 60 minutes. \lstinline+coreutils+ average 434 LOC. Our corpus
averages 2408 LOC, a 5-fold linear increase. Although Klee does not scale
linearly, we increased the time budget 5-fold to five hours, for both Klee and \Ourtool. 

\autoref{tab:unreachableBugs} reports our results: \Ourtool catches two bugs that Klee cannot catch: the bugs in \lstinline+ncompress+ and \lstinline+man+; both Klee and \Ourtool catch the bug in \lstinline+polymorph+. Neither tool catches the bug in \lstinline+gzip+. Not catching a bug means that the tool reaches the five hours timeout without generating a test case to reveal the bug in our benchmark~\citep{lu2005bugbench}. Our results show that \Ourtool reaches bugs previously unreachable for Klee in two cases out of four.

\subsection{Domain Restriction}
\label{subsec:outinth}

\Ourtool restricts the domain of operators involving \type to a subset of the initial supported values, the one that $G$ contains, to make them simpler:
\Ourtool rewrites constraints involving types in \type into the theory of
equality over integers.
We report the number of constraints on which \lstinline+STP+ times out: How many SMT queries return ``unknown'' before and after the application of \Ourtool?

When running Klee on all  coreutils, STP time-outs on 354
constraints. After indexifying the corpus, this number drops to 0.  In 53 out of the
89 coreutils programs, Klee times out before finishing;  only 6 programs time
out for \Ourproject. These results show \Ourtool generates indexed programs
that produce simpler constraints than Klee does when run directly on the
original program.

We then divided the 89 coreutils programs into 3 categories: projects with at least 5\% of the expressions indexed after applying \Ourtool; projects with at least 10\% of expressions indexed; and projects with at least 15\%  of expressions indexed. These 3 categories overlap, as we wanted to assess the performance of \Ourtool as a function of the percentage of indexed expressions. There were no important differences in the performance of \Ourtool between these three groups. \Ourtool is not dependent on the proportion of program expressions that  it converts, but rather on how much of the code in the program the paths that become reachable due to indexing make available for symex.

\subsection{\textsc{Indexify}ing Floats}
\label{subsec:floats}

To assess how well \Ourtool performs when indexing data types other than
strings,  we indexed the type \lstinline+float+ in our second benchmark:
fdlibm53. For $B_+$ we used the intersection between the functions involving
\lstinline+float+s in fdlibm53 and the ones in \lstinline+math.h+. We fixed  $k
= 3$, as the third quartile value of $k$, the length of chains of floating
point operators in our corpus (\autoref{fig:kreal}).\looseness=-1

\begin{wraptable}{l}{4.1cm}
\caption{Overall results of our experiments with  \Ourproject(float) and  $K=3$.}
\label{tab:resultsFloat} 
\small
\centering
\begin{tabular}{r r r}
\toprule 
\multicolumn{1}{c}{\textbf{Metric}} & \multicolumn{1}{c}{\textbf{$B_+^\circ$}}  & \multicolumn{1}{c}{\textbf{Klee}} \\
\midrule 
Time(min) & 10.45 &  0.84 \\
Memory(MB) & 71.98 & 6.85 \\
ICov(\%) & 83.35 & 50.91 \\
BCov(\%) & 71.56 & 34.45\\
\bottomrule 
\end{tabular}
\end{wraptable}

\autoref{tab:resultsFloat} shows \Ourproject's general effectiveness. \Ourtool
increased instruction coverage 33\% (row ``ICov'') and branch coverage 37\%
(row ``'BCov').  \Ourtool consumes ten times more memory than Klee (row
``Memory''). \Ourtool's increase in the run time from $0.84$ seconds in the case of Klee to $10.45$ seconds. This increase has two causes. The first is a very small execution times for fdlibm53.  Running \Ourtool has an additional runtime cost because of:  building the $\mathit{IOT}$s, changing the target type to indexes in the LLVM IR, and replacing the calls to the functions to be indexed, with their corresponding indexed versions. When the runtime to symex the initial program is big enough, the runtime reduction that our simpler constraints generate is bigger than the computational cost for \Ourtool. When the runtime to symex the initial program is very small, as is the case for our float benchmark, the preprocessing runtime to indexify the program is bigger than runtime reduction that the simpler constraints cause.  The second is the fact that more of the indexed program's constraints are solvable allows Klee to go further down paths, increasing coverage but incurring state space explosion.\looseness=-1

\subsection{The Size of Indexified Programs }

Indexification encodes the $\mathit{IOT}$s in the  IR representation of $P$. Although the disk space is cheap, it is important that the size of indexified programs are manageable in practice.

We used \lstinline+wc -l+ to report the LOCs of the  IR before and after indexification for coreutils. The total size in text lines for the initial coreutils is 1,731,007 lines; the total size for the indexified coreutils is 3,735,435. In the LLVM IR file format\footnote{\url{http://releases.llvm.org/2.7/docs/LangRef.html}} every instruction appears on a new line. Thus, we considered the newlines in LLVM IR as a good proxy for the number of instructions in that program.

Indexification increased the total size of the IR  for coreutils by 215\%. Although in the worst case,  the size of the $\mathit{IOT}$ grows exponentially in the average arity of the operators, this increase is manageable in practice. We speculate two reasons for this in coreutils. The programs in coreutils do not use all the string operators in \lstinline+string.h+: for example, the Unix tool \lstinline+yes+ does not use any string operators. \Ourtool adds additional code only for the $\mathit{IOT}$s. We construct an $\mathit{IOT}$ only for an operator that we index.  Second, \Ourtool encodes $\mathit{IOT}$ as a sequence of \lstinline+if+ statements, one per value in the garden.  LLVM translates each branch of an \lstinline+if+ statement into only 2 lines of IR: the predicate on one line, and the label for the true branch.\

\subsection{Number of Clauses}
\label{sec:clauses}

The number of clauses in queries affects the performance of symbolic execution. We compared the numbers of clauses between each original program and its indexified version. We ran Klee, and \Ourtool with the option: \lstinline+--use-query-log solver:all,pc+. This reports all the queries sent to the underlying solver, after Klee's internal optimisations.

After the application of \Ourtool, symbolically executing indexed programs generated 2,920,246,627 clauses in total, while the unindexed programs generated 997,389,929. Klee on indexed programs sent 44,138,715 clauses to STP, while it sent 
401,289,490 on unindexed programs.

\Ourtool generated three times as many constraints as Klee because of $\mathit{IOT}$s are encoded into queries sent to the solver. Klee's optimisations include simplifications that remove infeasible clauses from the queries, prior to sending them to the solver. These optimisations hugely benefit \Ourtool, removing irrelevant $\mathit{IOT}$ entries. After the optimisations, indexed programs generates $\frac{1}{10}$ the constraints that unindexed programs do. \Ourproject is very effective at simplifying constraints and Klee's solver chain is very effective at dealing with the simple constraints that \Ourproject produces.\looseness=-1

\section{Related work}
\label{sec:relwork}

Symbolic execution (symex) binds symbols to variables during execution.  When
it traverses a path, it constructs a path condition that define inputs
(including environmental interactions) that cause the program to take that
path.  Symex has some well known limitations~\citep{cadar2013symbolic} including
path explosion, coping with external code, and out-of-theory constraints.

Harman \etal~\citep{harman2004testability} introduced the concept of testability
transformations, \emph{source-to-source} program transformations whose goal is
to improve test data generation.  Following Harman \etal, Cadar speculated
that program transformations might improve the scalability of symbolic
execution in a position paper~\citep{cadar2015targeted}.  \Ourproject realizes,
in theory and practice, such a program transformation, rewriting a program to
allow symbolic execution to cover portions of the program's state space that
current symex engines cannot efficiently reach, as we show in \autoref{subsec:unique}. \Ourproject transforms expressions in a program that produce
\emph{out-of-theory} constraints into expressions that produce \emph{in-theory}
constraints.  The transformed (indexed)
program underapproximates the input program's behaviour:  it is precise 
over every component of the program's state it binds to a value in $G$ 
and reifies its ignorance of component's actual value when it binds $\bot$ 
to that component.

\subsection{Handling Unknown via Concretization}

Solver unknowns bedevil Symex. One way to handle them is to resort to concrete execution. \emph{Dynamic Symbolic Execution} (DSE)~\citep{cadar2008klee} does this by lazily concretizing the subset of the symbolic state for which the solver return unknown. Concolic testing~\citep{godefroid:dart, sen:cute, marinescu2012make} or \emph{white-box fuzzing}~\citep{godefroid2008automated}  is an extension to DSE
 that initially follows the path that a concrete input executes. Further, concolic testing searches a neighbourhood
around the path executed under its concrete input by negating the values of the current branch conditions from the current followed path. First concolic testing flips the closest branch to the end of execution and then, it continues to do so upwards to the entry point of the program. To flip the path condition, concolic testing uses an SMT solver. 

Whenever reaching a constraint that the solver cannot solve, or for which the solver times out, the concretization methods, such as DSE and concolic testing, get a concrete value that can be either random, or obtained under different heuristics~\citep{marinescu2012make}. In contrast \Ourtool keeps the value of the unsolvable constraint symbolic, but restricted to the garden.  This allows \Ourtool to obtain higher code coverage, by considering more paths than the concretization methods, as our direct comparison with Klee~\citep{cadar2008klee} (DSE) and  Zesti~\citep{marinescu2012make} (concolic testing) shows in \autoref{sec:B}. Additionally, \Ourtool does not require concrete inputs from the user, as concolic testing does: \Ourtool's builder automatically builds the set of values that it will consider in the analysis.\looseness=-1

\Ourproject generalizes the concretization techniques:  our garden  $G$ allows us to reason
about a finite set of values from $G$, instead of a single concrete value, as in the case of the concretization methods. 
Like concolic testing, \Ourtool generates test inputs from an initial set of values.
For \Ourtool, this initial set is the set of seeds harvested from constants;
for concolic testing and white-box fuzzing, the initial set is the input test suite.  \Ourproject and concolic testing generate these inputs
differently.   Concolic testing uses an SMT solver to negate the path
conditions of the initial inputs; \Ourproject constructs its garden using
repeated applications of builder functions.  Our intuition is that the
constants in a program are discrete points in the state space of the program: the constants appear in the predicates of the decisions points in the program.  By considering these points in $G$, we explore the behaviour of the program on the true branch of the predicate, the discrete point in the state space of the program. For
example, in the predicate \lstinline+if (strstr(S,"foo"))+, we would like to
consider the value \lstinline+foo+ for \lstinline+S+. Under \Ourproject
\lstinline+foo+ will certainly be considered, as \Ourproject will automatically
harvest into its seed set.  Concolic testing tends to generate inputs that
follow paths in the program near one of its initial inputs. \Ourtool, in
contrast, is not tethered  to a set of initial paths, so it can cover widely
varying program paths.

\subsection{Constraint Encoding}

During its execution, a program under Symex produces constraints in the different domains of its data types, such as constraints over strings, or floats. Before calling the solver, a Symex engine needs to encode the constraints in a language that the solver understands. Some of these constraints are difficult for the solver: the literature provide us a couple of constraint encoding mechanism to cope with the difficult kinds of constraints, in particular constraints over string, or floating point values.\looseness=-1 

\boldpara{Strings}
Quine proved that the first-order theory of string equations is
undecidable~\citep{quine1946concatenation}. Since then different techniques have
emerged implementing decision procedures over fragments of string theory. All
these approaches have limitations: either they do not scale; they support a
fragment of string theory that is not well-aligned with string expressions
developers write;  they require fixed length strings; or they require a maximum length and loop through all possible lengths up to that maximum.  There are three main categories of string solvers: automata,
bit-vector, and word-based.\looseness=-1 

Automata-based solvers~\citep{veanes2010symbolic} use regular languages or
context-free grammars to encode the string constraints. The idea is to
construct a finite-state automata that accepts all the strings that satisfy the
path conditions in a program.  Building this automata requires handcrafted building algorithms for the set of string operations that we intend to support and the set of values that the program accepts as inputs~\citep{hooimeijer2009decision}.  When a new
string constraint is added to the path condition, these approaches refine the
automaton to not accept the strings that violate the newly added constraint. The refinement process is automatic: we remove from the set of values that the automata accepts the ones that are not valid for the new constraint. Infeasible paths construct automata that do not accept any strings.  For string
constraints, the automaton becomes the solver.  Automata-based string solvers
tend not do not combine strings with other data types~\citep{li2013pass}, as
combining string automatas with other data types and operations over them, requires handcrafted initial automatas for the particular data types and operations that the program under analysis uses in conjunction with the string operations.  Yu
\etal~\citep{yu2009symbolic} tackles the problem of
using an automaton to handle both string and integer constraints, but no other
data types. Within the bounds of the values it handles, \Ourproject combines
different data types, including strings, by automatically translating them into
the theory of integer equality up to its garden, the finite set of values over
it is defined.

Bit-vector based symex engines convert string constraints into the domain of
bit-vectors~\citep{ganesh2007decision}.  The bit-vector solvers require a maximum string
length and lack scalability. Specifying a maximum length per each query that symex sends to the solver would require comprehensive annotations, so symex engines use a global maximum.   In general, string length is domain-specific and specifying a maximum length for an entire program is problematic: for
example, a database query might be hundreds of characters long, while a command
line flag can occupy only one character. The string length
restriction means that users must specify a single length for all strings in
the program.  When too big, this length slows the solver; when too small, it
limits the analysis to small strings.  Thus, these engines typically execute a
program over different length strings~\citep{ganesh2007decision}.  The
bit-vector encoding of values causes exponential blow up in model size, $2^n$
where $n$ is the length of a bit-vector, since each bit becomes a propositional
variable for the underlying SAT solver and hampers scalability. For a
restricted set of values, \Ourproject encodes constraints into table lookups,
not bit-vectors.  Running Klee on
programs transformed by \Ourtool covers more than twice the number of branches
that Klee manages on the unindexed programs (\autoref{sec:res}).  While Klee still
encodes the values in an indexed program's queries into bit-vectors, these are
short bit-vector encodings of integers, not long bit-vector encodings of
strings.\looseness=-1

Word-based string solvers define a subtheory that uses rewriting rules and
axioms tailored to a fragment of string theory. Word-based string solvers
escape Quine's result by not handling all string expressions.
CVC4~\citep{liang2014dpll} and S3~\citep{trinh2014s3} support constraints over
unbounded strings restricted only to length and regular expression membership
operators.  Z3-STR~\citep{zheng2013z3} supports unbounded strings together with
the concatenation, string equality, substring, replace, and length operators.
\Ourproject does not constrain the syntax of string expressions;  instead, it
constrains their values.  To do so, it finitizes string operators, restricting
their definition to a finite portion of their domain and range.  As
\autoref{sec:imp} describes, \Ourproject converts its finitized string
expressions into constraints over integer equality, then forwards them to an
SMT solver.

\boldpara{Floating Point} Bit-blasting is the most widely used technique for solving floating-point
constraints. Bit-blasting converts floats into bit-vectors and encodes floating
point operations into formulae over these bit-bectors. Bit-blasting
floating-point constraints generates  formulaes that require a huge number of variables. For example, Brillout \etal~\citep{brillout2009mixed} showed than when using a precision of only 5 (mantissa width) addition or subtraction over floating point variables requires a total of 1035 propositional variables  to be encoded in bit-vector theory. In a similar scenario, multiplication or division requires a total of 1048 propositional variables. Because of this, bit-blasting for floating point constraints
does not scale to large programs~\citep{darulova2014sound}.\looseness=-1

Testing is also used to solve floating point constraints. Microsoft's Pex symex engine can use the FloPSy~\citep{lakhotia2010flopsy} floating point solver.  FloPSy transforms  floating point (in)equalities into objective functions. For example, the predicate \lstinline+if((Math.Log(a) == b)+ becomes the objective function $| \mathit{Math.Log}(a) - b | $ for which we want to find values of $a$ and $b$ that yield $0$. FloPSy uses hill climbing~\citep{clarke1976system} to find values 
for $a$ and $b$. Fu \etal proposes Mathematical Execution (ME)~\citep{fu2016mathematical}. They capture the testing objective through a function that is minimised via mathematical optimisation to achieve the testing objective. Given a program under test $P$, they derive another program $P_R$ called representing function. $P_R$ represents how far an input $x \in \text{ dom}(P)$ is from reaching the set $x | P(x) \text{ is wrong}$. $P_R$ returns non-negative results and $P_R$ is the distance between the current input and an error triggering input. Further, they minimise $P_R$. Souza \etal~\citep{souza2011coral} integrate CORAL, a meta-heuristic solver designed for mathematical constraints, into PathFinder~\citep{visser2004test}. 

Klee-FP~\citep{collingbourne2011symbolic} and  Klee-CL~\citep{collingbourne2014symbolic} replaces floating-point instructions with uninterpreted functions. Klee-CL and Klee-FP apply a set of canonizing rewritings and further do a syntactic match of floating-point expressions trees. Both Klee-CL and Klee-FP solve floating point constraints by proving that them are equivalent (or not) with an integer only version of the program. They do this by matching the equivalent expression trees.  Thus, the main limitation is the requirement of having the both version of the programs available: the floating point and the integer implementation. \Ourtool does not require any existing integer implementation.

Other approaches use constraint
programming~\citep{michel2001solving} to soundly remove intervals of float
numbers that cannot make the path condition true. Botella
\etal~\citep{botella2006symbolic} solve floating point constraints using
interval propagation.  Interval propagation tries to contract the interval
domains of float variables without removing any value that might make the
constraint true. These approaches are either imprecise  or do not scale. \Ourtool translates floating point arithmetic into
equality theory over integers that is solvable in polynomial
time~\citep{barrett2009satisfiability}.

\subsection{The State Of the Art in Symbolic Execution} 

Significant research has tackled the problem of applying symex to real world
programs~\citep{cadar2013symbolic}.
\Ourproject complements this work: \Ourtool is a program transformation that is applied prior to symex; and thus can be applied in conjunction with any technique that improves symex. 

Some approaches aim to improve a symex engine's interactions with the constraint solver.
Klee~\citep{cadar2008klee} and Green~\citep{visser2012green} cache solved
constraints. Lazy initialization delays the concretization of
symbolic memory thereby avoiding the concretization of states that 
additional constraints make infeasible~\citep{rosner2015bliss}. 
Memoized symex~\citep{yang2012memoized} and directed incremental symbolic execution~\citep{person2011directed} reuse query results across different symex runs.\looseness=-1

Other approaches improve the scalability of Symex by mitigating the path explosion problem. Veritesting~/cite{avgerinos2014enhancing} proposes a path merging technique that reduces the number of paths being considered as a result of reasoning about the multiple merged paths simultaneously. MultiSE~\citep{sen2015multise} perform symbolic execution per method, rather than per the entire input program. MultiSE merges different symex paths into a value summary, or a conditional execution state. Further, we can symex each method in a program starting from its value summary, rather than from the program's entry point.

For an unsolvable constraint, symex concretizes its variables, causing incompleteness: it cannot reason on the rest of the state space. Pasareanu \etal~ \citep{puasuareanu2011symbolic} delay concretization to limit the incompleteness. They divide the clauses into simple and complex ones. When a simple clause is unsatisfiable, the entire PC becomes unsatisfiable. This can avoid reasoning about the complex clauses.  Khurshid \etal~\citep{khurshid2003generalized} concretizes objects only when they need to access them.\looseness=-1

A recent study~\citep{dong2015studying} show that 33 optimization flags in \lstinline+LLVM+ decrease symex's coverage on \lstinline+coreutils+. Overify~\citep{wagner2013overify} proposes a set of compiler optimisations to speed symex. Sharma \etal~\citep{sharma2014exploiting} exploit these results and introduce undefined behaviour to trigger various compiler optimisations that speed up symex~\citep{cadar2015targeted}.  Under-Constrained Symex~\citep{ramos2015under} operates on each function in a program individually. Abstract subsumption~\citep{anand2006symbolic} checks for symbolic states that subsume other ones and remove the subsumed ones. Ariadne transforms numerical programs to explicitly check exception triggering conditions in the case of floats~\citep{barr2013automatic}. As \Ourproject is a program transformation step prior to symex, we can take advantage of these optimisations by running \Ourproject in conjunction with them.\looseness=-1

\section{Conclusion}

We introduced indexification, a novel technique that rewrites a program to
constrain its behaviour to a subset of its original state space.  Over this
restricted space, the rewritten program under-approximates the original; its
symbolic execution generates tractable constraints.  We realized indexification
in \Ourtool and show that it automatically harvests program constants to define
restricted space permitting symex to explore paths and find bugs that other
symbolic execution techniques do not reach.

\newpage
\balance

\bibliographystyle{ACM-Reference-Format}

\bibliography{slice}


\begin{thebibliography}{56}


\ifx \showCODEN    \undefined \def \showCODEN     #1{\unskip}     \fi
\ifx \showDOI      \undefined \def \showDOI       #1{#1}\fi
\ifx \showISBNx    \undefined \def \showISBNx     #1{\unskip}     \fi
\ifx \showISBNxiii \undefined \def \showISBNxiii  #1{\unskip}     \fi
\ifx \showISSN     \undefined \def \showISSN      #1{\unskip}     \fi
\ifx \showLCCN     \undefined \def \showLCCN      #1{\unskip}     \fi
\ifx \shownote     \undefined \def \shownote      #1{#1}          \fi
\ifx \showarticletitle \undefined \def \showarticletitle #1{#1}   \fi
\ifx \showURL      \undefined \def \showURL       {\relax}        \fi
\providecommand\bibfield[2]{#2}
\providecommand\bibinfo[2]{#2}
\providecommand\natexlab[1]{#1}
\providecommand\showeprint[2][]{arXiv:#2}

\bibitem[\protect\citeauthoryear{Anand, P{\u{a}}s{\u{a}}reanu, and
  Visser}{Anand et~al\mbox{.}}{2006}]%
        {anand2006symbolic}
\bibfield{author}{\bibinfo{person}{Saswat Anand}, \bibinfo{person}{Corina~S
  P{\u{a}}s{\u{a}}reanu}, {and} \bibinfo{person}{Willem Visser}.}
  \bibinfo{year}{2006}\natexlab{}.
\newblock \showarticletitle{Symbolic execution with abstract subsumption
  checking}. In \bibinfo{booktitle}{\emph{International SPIN Workshop on Model
  Checking of Software}}. Springer, \bibinfo{pages}{163--181}.
\newblock


\bibitem[\protect\citeauthoryear{Barr, Vo, Le, and Su}{Barr
  et~al\mbox{.}}{2013}]%
        {barr2013automatic}
\bibfield{author}{\bibinfo{person}{Earl~T Barr}, \bibinfo{person}{Thanh Vo},
  \bibinfo{person}{Vu Le}, {and} \bibinfo{person}{Zhendong Su}.}
  \bibinfo{year}{2013}\natexlab{}.
\newblock \showarticletitle{Automatic detection of floating-point exceptions}.
  In \bibinfo{booktitle}{\emph{ACM SIGPLAN Notices}},
  Vol.~\bibinfo{volume}{48}. \bibinfo{publisher}{ACM},
  \bibinfo{pages}{549--560}.
\newblock


\bibitem[\protect\citeauthoryear{Barrett, Sebastiani, Seshia, and
  Tinelli}{Barrett et~al\mbox{.}}{2009}]%
        {barrett2009satisfiability}
\bibfield{author}{\bibinfo{person}{Clark~W Barrett}, \bibinfo{person}{Roberto
  Sebastiani}, \bibinfo{person}{Sanjit~A Seshia}, {and} \bibinfo{person}{Cesare
  Tinelli}.} \bibinfo{year}{2009}\natexlab{}.
\newblock \showarticletitle{Satisfiability Modulo Theories.}
\newblock \bibinfo{journal}{\emph{Handbook of satisfiability}}
  \bibinfo{volume}{185} (\bibinfo{year}{2009}), \bibinfo{pages}{825--885}.
\newblock


\bibitem[\protect\citeauthoryear{Bj{\o}rner, Tillmann, and Voronkov}{Bj{\o}rner
  et~al\mbox{.}}{2009}]%
        {bjorner2009path}
\bibfield{author}{\bibinfo{person}{Nikolaj Bj{\o}rner},
  \bibinfo{person}{Nikolai Tillmann}, {and} \bibinfo{person}{Andrei Voronkov}.}
  \bibinfo{year}{2009}\natexlab{}.
\newblock \showarticletitle{Path feasibility analysis for string-manipulating
  programs}.
\newblock In \bibinfo{booktitle}{\emph{Tools and Algorithms for the
  Construction and Analysis of Systems}}. \bibinfo{publisher}{Springer},
  \bibinfo{pages}{307--321}.
\newblock


\bibitem[\protect\citeauthoryear{Botella, Gotlieb, and Michel}{Botella
  et~al\mbox{.}}{2006}]%
        {botella2006symbolic}
\bibfield{author}{\bibinfo{person}{Bernard Botella}, \bibinfo{person}{Arnaud
  Gotlieb}, {and} \bibinfo{person}{Claude Michel}.}
  \bibinfo{year}{2006}\natexlab{}.
\newblock \showarticletitle{Symbolic execution of floating-point computations}.
\newblock \bibinfo{journal}{\emph{Software Testing, Verification and
  Reliability}} \bibinfo{volume}{16}, \bibinfo{number}{2}
  (\bibinfo{year}{2006}), \bibinfo{pages}{97--121}.
\newblock


\bibitem[\protect\citeauthoryear{Brillout, Kroening, and Wahl}{Brillout
  et~al\mbox{.}}{2009}]%
        {brillout2009mixed}
\bibfield{author}{\bibinfo{person}{Angelo Brillout}, \bibinfo{person}{Daniel
  Kroening}, {and} \bibinfo{person}{Thomas Wahl}.}
  \bibinfo{year}{2009}\natexlab{}.
\newblock \showarticletitle{Mixed abstractions for floating-point arithmetic}.
  In \bibinfo{booktitle}{\emph{Formal Methods in Computer-Aided Design, 2009.
  FMCAD 2009}}. IEEE, \bibinfo{pages}{69--76}.
\newblock


\bibitem[\protect\citeauthoryear{Cadar}{Cadar}{2015}]%
        {cadar2015targeted}
\bibfield{author}{\bibinfo{person}{Cristian Cadar}.}
  \bibinfo{year}{2015}\natexlab{}.
\newblock \showarticletitle{Targeted program transformations for symbolic
  execution}. In \bibinfo{booktitle}{\emph{Proceedings of the 2015 10th Joint
  Meeting on Foundations of Software Engineering}}. ACM,
  \bibinfo{pages}{906--909}.
\newblock


\bibitem[\protect\citeauthoryear{Cadar, Dunbar, Engler, et~al\mbox{.}}{Cadar
  et~al\mbox{.}}{2008}]%
        {cadar2008klee}
\bibfield{author}{\bibinfo{person}{Cristian Cadar}, \bibinfo{person}{Daniel
  Dunbar}, \bibinfo{person}{Dawson~R Engler}, {et~al\mbox{.}}}
  \bibinfo{year}{2008}\natexlab{}.
\newblock \showarticletitle{KLEE: Unassisted and Automatic Generation of
  High-Coverage Tests for Complex Systems Programs.}. In
  \bibinfo{booktitle}{\emph{OSDI}}, Vol.~\bibinfo{volume}{8}.
  \bibinfo{pages}{209--224}.
\newblock


\bibitem[\protect\citeauthoryear{Cadar and Sen}{Cadar and Sen}{2013}]%
        {cadar2013symbolic}
\bibfield{author}{\bibinfo{person}{Cristian Cadar} {and}
  \bibinfo{person}{Koushik Sen}.} \bibinfo{year}{2013}\natexlab{}.
\newblock \showarticletitle{Symbolic execution for software testing: three
  decades later}.
\newblock \bibinfo{journal}{\emph{Commun. ACM}} \bibinfo{volume}{56},
  \bibinfo{number}{2} (\bibinfo{year}{2013}), \bibinfo{pages}{82--90}.
\newblock


\bibitem[\protect\citeauthoryear{Chen, Zhang, Guo, Li, and Wu}{Chen
  et~al\mbox{.}}{2013}]%
        {chen2013state}
\bibfield{author}{\bibinfo{person}{Ting Chen}, \bibinfo{person}{Xiao-song
  Zhang}, \bibinfo{person}{Shi-ze Guo}, \bibinfo{person}{Hong-yuan Li}, {and}
  \bibinfo{person}{Yue Wu}.} \bibinfo{year}{2013}\natexlab{}.
\newblock \showarticletitle{State of the art: Dynamic symbolic execution for
  automated test generation}.
\newblock \bibinfo{journal}{\emph{Future Generation Computer Systems}}
  \bibinfo{volume}{29}, \bibinfo{number}{7} (\bibinfo{year}{2013}),
  \bibinfo{pages}{1758--1773}.
\newblock


\bibitem[\protect\citeauthoryear{Clarke}{Clarke}{1976}]%
        {clarke1976system}
\bibfield{author}{\bibinfo{person}{Lori~A. Clarke}.}
  \bibinfo{year}{1976}\natexlab{}.
\newblock \showarticletitle{A system to generate test data and symbolically
  execute programs}.
\newblock \bibinfo{journal}{\emph{IEEE Transactions on software engineering}}
  \bibinfo{number}{3} (\bibinfo{year}{1976}), \bibinfo{pages}{215--222}.
\newblock


\bibitem[\protect\citeauthoryear{Collingbourne, Cadar, and Kelly}{Collingbourne
  et~al\mbox{.}}{2011}]%
        {collingbourne2011symbolic}
\bibfield{author}{\bibinfo{person}{Peter Collingbourne},
  \bibinfo{person}{Cristian Cadar}, {and} \bibinfo{person}{Paul~HJ Kelly}.}
  \bibinfo{year}{2011}\natexlab{}.
\newblock \showarticletitle{Symbolic crosschecking of floating-point and SIMD
  code}. In \bibinfo{booktitle}{\emph{Proceedings of the sixth conference on
  Computer systems}}. ACM, \bibinfo{pages}{315--328}.
\newblock


\bibitem[\protect\citeauthoryear{Collingbourne, Cadar, and Kelly}{Collingbourne
  et~al\mbox{.}}{2014}]%
        {collingbourne2014symbolic}
\bibfield{author}{\bibinfo{person}{Peter Collingbourne},
  \bibinfo{person}{Cristian Cadar}, {and} \bibinfo{person}{Paul~HJ Kelly}.}
  \bibinfo{year}{2014}\natexlab{}.
\newblock \showarticletitle{Symbolic crosschecking of data-parallel
  floating-point code}.
\newblock \bibinfo{journal}{\emph{IEEE Transactions on Software Engineering}}
  \bibinfo{volume}{40}, \bibinfo{number}{7} (\bibinfo{year}{2014}),
  \bibinfo{pages}{710--737}.
\newblock


\bibitem[\protect\citeauthoryear{Darulova and Kuncak}{Darulova and
  Kuncak}{2014}]%
        {darulova2014sound}
\bibfield{author}{\bibinfo{person}{Eva Darulova} {and} \bibinfo{person}{Viktor
  Kuncak}.} \bibinfo{year}{2014}\natexlab{}.
\newblock \showarticletitle{Sound compilation of reals}.
\newblock \bibinfo{journal}{\emph{Acm Sigplan Notices}} \bibinfo{volume}{49},
  \bibinfo{number}{1} (\bibinfo{year}{2014}), \bibinfo{pages}{235--248}.
\newblock


\bibitem[\protect\citeauthoryear{Developers}{Developers}{2016a}]%
        {llvm}
\bibfield{author}{\bibinfo{person}{LLVM Developers}.}
  \bibinfo{year}{2016}\natexlab{a}.
\newblock \bibinfo{title}{LLVM}.
\newblock \bibinfo{howpublished}{\url{http://llvm.org/}}.
  (\bibinfo{year}{2016}).
\newblock
\newblock
\shownote{Accessed: 2016-07-02.}


\bibitem[\protect\citeauthoryear{Developers}{Developers}{2016b}]%
        {reactOSWeb}
\bibfield{author}{\bibinfo{person}{ReactOS Developers}.}
  \bibinfo{year}{2016}\natexlab{b}.
\newblock \bibinfo{title}{ReactOS}.
\newblock \bibinfo{howpublished}{\url{http://www.reactos.org/}}.
  (\bibinfo{year}{2016}).
\newblock
\newblock
\shownote{Accessed: 2016-07-02.}


\bibitem[\protect\citeauthoryear{Dillig, Dillig, and Aiken}{Dillig
  et~al\mbox{.}}{2010}]%
        {dillig2010small}
\bibfield{author}{\bibinfo{person}{Isil Dillig}, \bibinfo{person}{Thomas
  Dillig}, {and} \bibinfo{person}{Alex Aiken}.}
  \bibinfo{year}{2010}\natexlab{}.
\newblock \showarticletitle{Small formulas for large programs: On-line
  constraint simplification in scalable static analysis}. In
  \bibinfo{booktitle}{\emph{International Static Analysis Symposium}}.
  Springer, \bibinfo{pages}{236--252}.
\newblock


\bibitem[\protect\citeauthoryear{Dong, Olivo, Zhang, and Khurshid}{Dong
  et~al\mbox{.}}{2015}]%
        {dong2015studying}
\bibfield{author}{\bibinfo{person}{Shiyu Dong}, \bibinfo{person}{Oswaldo
  Olivo}, \bibinfo{person}{Lingming Zhang}, {and} \bibinfo{person}{Sarfraz
  Khurshid}.} \bibinfo{year}{2015}\natexlab{}.
\newblock \showarticletitle{Studying the influence of standard compiler
  optimizations on symbolic execution}. In \bibinfo{booktitle}{\emph{Software
  Reliability Engineering (ISSRE), 2015 IEEE 26th International Symposium on}}.
  IEEE, \bibinfo{pages}{205--215}.
\newblock


\bibitem[\protect\citeauthoryear{Fu and Su}{Fu and Su}{2016}]%
        {fu2016mathematical}
\bibfield{author}{\bibinfo{person}{Zhoulai Fu} {and} \bibinfo{person}{Zhendong
  Su}.} \bibinfo{year}{2016}\natexlab{}.
\newblock \showarticletitle{Mathematical Execution: A Unified Approach for
  Testing Numerical Code}.
\newblock \bibinfo{journal}{\emph{arXiv preprint arXiv:1610.01133}}
  (\bibinfo{year}{2016}).
\newblock


\bibitem[\protect\citeauthoryear{Ganesh and Dill}{Ganesh and Dill}{2007}]%
        {ganesh2007decision}
\bibfield{author}{\bibinfo{person}{Vijay Ganesh} {and} \bibinfo{person}{David~L
  Dill}.} \bibinfo{year}{2007}\natexlab{}.
\newblock \showarticletitle{A decision procedure for bit-vectors and arrays}.
  In \bibinfo{booktitle}{\emph{International Conference on Computer Aided
  Verification}}. Springer, \bibinfo{pages}{519--531}.
\newblock


\bibitem[\protect\citeauthoryear{Gligoric, Groce, Zhang, Sharma, Alipour, and
  Marinov}{Gligoric et~al\mbox{.}}{2013}]%
        {gligoric:comparing}
\bibfield{author}{\bibinfo{person}{Milos Gligoric}, \bibinfo{person}{Alex
  Groce}, \bibinfo{person}{Chaoqiang Zhang}, \bibinfo{person}{Rohan Sharma},
  \bibinfo{person}{Mohammad~Amin Alipour}, {and} \bibinfo{person}{Darko
  Marinov}.} \bibinfo{year}{2013}\natexlab{}.
\newblock \showarticletitle{Comparing non-adequate test suites using coverage
  criteria}. In \bibinfo{booktitle}{\emph{International Symposium on Software
  Testing and Analysis {(ISSTA 2013)}}},
  \bibfield{editor}{\bibinfo{person}{Mauro Pezz{\`e}} {and}
  \bibinfo{person}{Mark Harman}} (Eds.). \bibinfo{publisher}{ACM},
  \bibinfo{address}{Lugano, Switzerland}, \bibinfo{pages}{302--313}.
\newblock
\showISBNx{978-1-4503-2159-4}


\bibitem[\protect\citeauthoryear{Godefroid, Klarlund, and Sen}{Godefroid
  et~al\mbox{.}}{2005}]%
        {godefroid:dart}
\bibfield{author}{\bibinfo{person}{Patrice Godefroid}, \bibinfo{person}{Nils
  Klarlund}, {and} \bibinfo{person}{Koushik Sen}.}
  \bibinfo{year}{2005}\natexlab{}.
\newblock \showarticletitle{{DART}: directed automated random testing}. In
  \bibinfo{booktitle}{\emph{Programming Language Design and Implementation
  ({PLDI 2005})}}, \bibfield{editor}{\bibinfo{person}{Vivek Sarkar} {and}
  \bibinfo{person}{Mary~W. Hall}} (Eds.). \bibinfo{publisher}{ACM},
  \bibinfo{pages}{213--223}.
\newblock
\showISBNx{1-59593-056-6}


\bibitem[\protect\citeauthoryear{Godefroid, Levin, Molnar,
  et~al\mbox{.}}{Godefroid et~al\mbox{.}}{2008}]%
        {godefroid2008automated}
\bibfield{author}{\bibinfo{person}{Patrice Godefroid},
  \bibinfo{person}{Michael~Y Levin}, \bibinfo{person}{David~A Molnar},
  {et~al\mbox{.}}} \bibinfo{year}{2008}\natexlab{}.
\newblock \showarticletitle{Automated Whitebox Fuzz Testing.}. In
  \bibinfo{booktitle}{\emph{NDSS}}, Vol.~\bibinfo{volume}{8}.
  \bibinfo{pages}{151--166}.
\newblock


\bibitem[\protect\citeauthoryear{Harman, Hu, Hierons, Wegener, Sthamer,
  Baresel, and Roper}{Harman et~al\mbox{.}}{2004}]%
        {harman2004testability}
\bibfield{author}{\bibinfo{person}{Mark Harman}, \bibinfo{person}{Lin Hu},
  \bibinfo{person}{Rob Hierons}, \bibinfo{person}{Joachim Wegener},
  \bibinfo{person}{Harmen Sthamer}, \bibinfo{person}{Andr{\'e} Baresel}, {and}
  \bibinfo{person}{Marc Roper}.} \bibinfo{year}{2004}\natexlab{}.
\newblock \showarticletitle{Testability transformation}.
\newblock \bibinfo{journal}{\emph{IEEE Transactions on Software Engineering}}
  \bibinfo{volume}{30}, \bibinfo{number}{1} (\bibinfo{year}{2004}),
  \bibinfo{pages}{3--16}.
\newblock


\bibitem[\protect\citeauthoryear{Hooimeijer and Weimer}{Hooimeijer and
  Weimer}{2009}]%
        {hooimeijer2009decision}
\bibfield{author}{\bibinfo{person}{Pieter Hooimeijer} {and}
  \bibinfo{person}{Westley Weimer}.} \bibinfo{year}{2009}\natexlab{}.
\newblock \showarticletitle{A decision procedure for subset constraints over
  regular languages}.
\newblock \bibinfo{journal}{\emph{ACM Sigplan Notices}} \bibinfo{volume}{44},
  \bibinfo{number}{6} (\bibinfo{year}{2009}), \bibinfo{pages}{188--198}.
\newblock


\bibitem[\protect\citeauthoryear{Jha, Seshia, and Limaye}{Jha
  et~al\mbox{.}}{2009}]%
        {jha2009computational}
\bibfield{author}{\bibinfo{person}{Susmit Jha}, \bibinfo{person}{Sanjit~A
  Seshia}, {and} \bibinfo{person}{Rhishikesh Limaye}.}
  \bibinfo{year}{2009}\natexlab{}.
\newblock \showarticletitle{On the computational complexity of satisfiability
  solving for string theories}.
\newblock \bibinfo{journal}{\emph{arXiv preprint arXiv:0903.2825}}
  (\bibinfo{year}{2009}).
\newblock


\bibitem[\protect\citeauthoryear{Khurshid, P{\u{a}}s{\u{a}}reanu, and
  Visser}{Khurshid et~al\mbox{.}}{2003}]%
        {khurshid2003generalized}
\bibfield{author}{\bibinfo{person}{Sarfraz Khurshid}, \bibinfo{person}{Corina~S
  P{\u{a}}s{\u{a}}reanu}, {and} \bibinfo{person}{Willem Visser}.}
  \bibinfo{year}{2003}\natexlab{}.
\newblock \showarticletitle{Generalized symbolic execution for model checking
  and testing}. In \bibinfo{booktitle}{\emph{International Conference on Tools
  and Algorithms for the Construction and Analysis of Systems}}. Springer,
  \bibinfo{pages}{553--568}.
\newblock


\bibitem[\protect\citeauthoryear{Kleene}{Kleene}{1951}]%
        {kleene1951representation}
\bibfield{author}{\bibinfo{person}{Stephen~Cole Kleene}.}
  \bibinfo{year}{1951}\natexlab{}.
\newblock \bibinfo{booktitle}{\emph{Representation of events in nerve nets and
  finite automata}}.
\newblock \bibinfo{type}{{T}echnical {R}eport}. \bibinfo{institution}{DTIC
  Document}.
\newblock


\bibitem[\protect\citeauthoryear{Kroening and Tautschnig}{Kroening and
  Tautschnig}{2014}]%
        {kroening2014cbmc}
\bibfield{author}{\bibinfo{person}{Daniel Kroening} {and}
  \bibinfo{person}{Michael Tautschnig}.} \bibinfo{year}{2014}\natexlab{}.
\newblock \showarticletitle{CBMC--C bounded model checker}. In
  \bibinfo{booktitle}{\emph{International Conference on Tools and Algorithms
  for the Construction and Analysis of Systems}}. Springer,
  \bibinfo{pages}{389--391}.
\newblock


\bibitem[\protect\citeauthoryear{Lakhotia, McMinn, and Harman}{Lakhotia
  et~al\mbox{.}}{2009}]%
        {kletal:taicpart09}
\bibfield{author}{\bibinfo{person}{Kiran Lakhotia}, \bibinfo{person}{Phil
  McMinn}, {and} \bibinfo{person}{Mark Harman}.}
  \bibinfo{year}{2009}\natexlab{}.
\newblock \showarticletitle{Automated Test Data Generation for Coverage:
  {H}aven't We Solved This Problem Yet?}. In \bibinfo{booktitle}{\emph{$4^{th}$
  {T}esting {A}cademia and {I}ndustry {C}onference --- {P}ractice {A}nd
  {R}esearch {T}echniques ({TAIC PART'09})}}. \bibinfo{address}{Windsor, UK},
  \bibinfo{pages}{95--104}.
\newblock


\bibitem[\protect\citeauthoryear{Lakhotia, Tillmann, Harman, and
  De~Halleux}{Lakhotia et~al\mbox{.}}{2010}]%
        {lakhotia2010flopsy}
\bibfield{author}{\bibinfo{person}{Kiran Lakhotia}, \bibinfo{person}{Nikolai
  Tillmann}, \bibinfo{person}{Mark Harman}, {and} \bibinfo{person}{Jonathan
  De~Halleux}.} \bibinfo{year}{2010}\natexlab{}.
\newblock \showarticletitle{Flopsy-search-based floating point constraint
  solving for symbolic execution}. In \bibinfo{booktitle}{\emph{IFIP
  International Conference on Testing Software and Systems}}. Springer,
  \bibinfo{pages}{142--157}.
\newblock


\bibitem[\protect\citeauthoryear{Li and Ghosh}{Li and Ghosh}{2013}]%
        {li2013pass}
\bibfield{author}{\bibinfo{person}{Guodong Li} {and} \bibinfo{person}{Indradeep
  Ghosh}.} \bibinfo{year}{2013}\natexlab{}.
\newblock \showarticletitle{PASS: String solving with parameterized array and
  interval automaton}. In \bibinfo{booktitle}{\emph{Haifa Verification
  Conference}}. Springer, \bibinfo{pages}{15--31}.
\newblock


\bibitem[\protect\citeauthoryear{Liang, Reynolds, Tinelli, Barrett, and
  Deters}{Liang et~al\mbox{.}}{2014}]%
        {liang2014dpll}
\bibfield{author}{\bibinfo{person}{Tianyi Liang}, \bibinfo{person}{Andrew
  Reynolds}, \bibinfo{person}{Cesare Tinelli}, \bibinfo{person}{Clark Barrett},
  {and} \bibinfo{person}{Morgan Deters}.} \bibinfo{year}{2014}\natexlab{}.
\newblock \showarticletitle{A DPLL (T) theory solver for a theory of strings
  and regular expressions}. In \bibinfo{booktitle}{\emph{International
  Conference on Computer Aided Verification}}. Springer,
  \bibinfo{pages}{646--662}.
\newblock


\bibitem[\protect\citeauthoryear{Lu, Li, Qin, Tan, Zhou, and Zhou}{Lu
  et~al\mbox{.}}{2005}]%
        {lu2005bugbench}
\bibfield{author}{\bibinfo{person}{Shan Lu}, \bibinfo{person}{Zhenmin Li},
  \bibinfo{person}{Feng Qin}, \bibinfo{person}{Lin Tan}, \bibinfo{person}{Pin
  Zhou}, {and} \bibinfo{person}{Yuanyuan Zhou}.}
  \bibinfo{year}{2005}\natexlab{}.
\newblock \showarticletitle{Bugbench: Benchmarks for evaluating bug detection
  tools}. In \bibinfo{booktitle}{\emph{Workshop on the evaluation of software
  defect detection tools}}, Vol.~\bibinfo{volume}{5}.
\newblock


\bibitem[\protect\citeauthoryear{Marinescu and Cadar}{Marinescu and
  Cadar}{2012}]%
        {marinescu2012make}
\bibfield{author}{\bibinfo{person}{Paul~Dan Marinescu} {and}
  \bibinfo{person}{Cristian Cadar}.} \bibinfo{year}{2012}\natexlab{}.
\newblock \showarticletitle{Make test-zesti: A symbolic execution solution for
  improving regression testing}. In \bibinfo{booktitle}{\emph{Proceedings of
  the 34th International Conference on Software Engineering}}. IEEE Press,
  \bibinfo{pages}{716--726}.
\newblock


\bibitem[\protect\citeauthoryear{Michel, Rueher, and Lebbah}{Michel
  et~al\mbox{.}}{2001}]%
        {michel2001solving}
\bibfield{author}{\bibinfo{person}{Claude Michel}, \bibinfo{person}{Michel
  Rueher}, {and} \bibinfo{person}{Yahia Lebbah}.}
  \bibinfo{year}{2001}\natexlab{}.
\newblock \showarticletitle{Solving constraints over floating-point numbers}.
  In \bibinfo{booktitle}{\emph{International Conference on Principles and
  Practice of Constraint Programming}}. Springer, \bibinfo{pages}{524--538}.
\newblock


\bibitem[\protect\citeauthoryear{P{\u{a}}s{\u{a}}reanu, Rungta, and
  Visser}{P{\u{a}}s{\u{a}}reanu et~al\mbox{.}}{2011}]%
        {puasuareanu2011symbolic}
\bibfield{author}{\bibinfo{person}{Corina~S P{\u{a}}s{\u{a}}reanu},
  \bibinfo{person}{Neha Rungta}, {and} \bibinfo{person}{Willem Visser}.}
  \bibinfo{year}{2011}\natexlab{}.
\newblock \showarticletitle{Symbolic execution with mixed concrete-symbolic
  solving}. In \bibinfo{booktitle}{\emph{Proceedings of the 2011 International
  Symposium on Software Testing and Analysis}}. ACM, \bibinfo{pages}{34--44}.
\newblock


\bibitem[\protect\citeauthoryear{Person, Yang, Rungta, and Khurshid}{Person
  et~al\mbox{.}}{2011}]%
        {person2011directed}
\bibfield{author}{\bibinfo{person}{Suzette Person}, \bibinfo{person}{Guowei
  Yang}, \bibinfo{person}{Neha Rungta}, {and} \bibinfo{person}{Sarfraz
  Khurshid}.} \bibinfo{year}{2011}\natexlab{}.
\newblock \showarticletitle{Directed incremental symbolic execution}. In
  \bibinfo{booktitle}{\emph{ACM SIGPLAN Notices}}, Vol.~\bibinfo{volume}{46}.
  ACM, \bibinfo{pages}{504--515}.
\newblock


\bibitem[\protect\citeauthoryear{Quine}{Quine}{1946}]%
        {quine1946concatenation}
\bibfield{author}{\bibinfo{person}{Willard~V Quine}.}
  \bibinfo{year}{1946}\natexlab{}.
\newblock \showarticletitle{Concatenation as a basis for arithmetic}.
\newblock \bibinfo{journal}{\emph{The Journal of Symbolic Logic}}
  \bibinfo{volume}{11}, \bibinfo{number}{04} (\bibinfo{year}{1946}),
  \bibinfo{pages}{105--114}.
\newblock


\bibitem[\protect\citeauthoryear{Ramos and Engler}{Ramos and Engler}{2015}]%
        {ramos2015under}
\bibfield{author}{\bibinfo{person}{David~A Ramos} {and}
  \bibinfo{person}{Dawson~R Engler}.} \bibinfo{year}{2015}\natexlab{}.
\newblock \showarticletitle{Under-Constrained Symbolic Execution: Correctness
  Checking for Real Code.}. In \bibinfo{booktitle}{\emph{USENIX Security}}.
  \bibinfo{pages}{49--64}.
\newblock


\bibitem[\protect\citeauthoryear{Rosen}{Rosen}{1973}]%
        {rosen1973tree}
\bibfield{author}{\bibinfo{person}{Barry~K Rosen}.}
  \bibinfo{year}{1973}\natexlab{}.
\newblock \showarticletitle{Tree-manipulating systems and Church-Rosser
  theorems}.
\newblock \bibinfo{journal}{\emph{Journal of the ACM (JACM)}}
  \bibinfo{volume}{20}, \bibinfo{number}{1} (\bibinfo{year}{1973}),
  \bibinfo{pages}{160--187}.
\newblock


\bibitem[\protect\citeauthoryear{Rosner, Geldenhuys, Aguirre, Visser, and
  Frias}{Rosner et~al\mbox{.}}{2015}]%
        {rosner2015bliss}
\bibfield{author}{\bibinfo{person}{Nicol{\'a}s Rosner}, \bibinfo{person}{Jaco
  Geldenhuys}, \bibinfo{person}{Nazareno~M Aguirre}, \bibinfo{person}{Willem
  Visser}, {and} \bibinfo{person}{Marcelo~F Frias}.}
  \bibinfo{year}{2015}\natexlab{}.
\newblock \showarticletitle{BLISS: Improved symbolic execution by bounded lazy
  initialization with sat support}.
\newblock \bibinfo{journal}{\emph{IEEE Transactions on Software Engineering}}
  \bibinfo{volume}{41}, \bibinfo{number}{7} (\bibinfo{year}{2015}),
  \bibinfo{pages}{639--660}.
\newblock


\bibitem[\protect\citeauthoryear{Sen, Marinov, and Agha}{Sen
  et~al\mbox{.}}{2005a}]%
        {sen2005cute}
\bibfield{author}{\bibinfo{person}{Koushik Sen}, \bibinfo{person}{Darko
  Marinov}, {and} \bibinfo{person}{Gul Agha}.}
  \bibinfo{year}{2005}\natexlab{a}.
\newblock \showarticletitle{CUTE: a concolic unit testing engine for C}. In
  \bibinfo{booktitle}{\emph{ACM SIGSOFT Software Engineering Notes}},
  Vol.~\bibinfo{volume}{30}. ACM, \bibinfo{pages}{263--272}.
\newblock


\bibitem[\protect\citeauthoryear{Sen, Marinov, and Agha}{Sen
  et~al\mbox{.}}{2005b}]%
        {sen:cute}
\bibfield{author}{\bibinfo{person}{Koushik Sen}, \bibinfo{person}{Darko
  Marinov}, {and} \bibinfo{person}{Gul Agha}.}
  \bibinfo{year}{2005}\natexlab{b}.
\newblock \showarticletitle{{CUTE}: a concolic unit testing engine for {C}}. In
  \bibinfo{booktitle}{\emph{$10^{th}$ European Software Engineering Conference
  and 13th {ACM} International Symposium on Foundations of Software Engineering
  ({ESEC/FSE} '05)}}, \bibfield{editor}{\bibinfo{person}{Michel Wermelinger}
  {and} \bibinfo{person}{Harald Gall}} (Eds.). \bibinfo{publisher}{ACM},
  \bibinfo{pages}{263--272}.
\newblock
\showISBNx{1-59593-014-0}


\bibitem[\protect\citeauthoryear{Sen, Necula, Gong, and Choi}{Sen
  et~al\mbox{.}}{2015}]%
        {sen2015multise}
\bibfield{author}{\bibinfo{person}{Koushik Sen}, \bibinfo{person}{George
  Necula}, \bibinfo{person}{Liang Gong}, {and} \bibinfo{person}{Wontae Choi}.}
  \bibinfo{year}{2015}\natexlab{}.
\newblock \showarticletitle{MultiSE: Multi-path symbolic execution using value
  summaries}. In \bibinfo{booktitle}{\emph{Proceedings of the 2015 10th Joint
  Meeting on Foundations of Software Engineering}}. ACM,
  \bibinfo{pages}{842--853}.
\newblock


\bibitem[\protect\citeauthoryear{Sharma}{Sharma}{2014}]%
        {sharma2014exploiting}
\bibfield{author}{\bibinfo{person}{Asankhaya Sharma}.}
  \bibinfo{year}{2014}\natexlab{}.
\newblock \showarticletitle{Exploiting undefined behaviors for efficient
  symbolic execution}. In \bibinfo{booktitle}{\emph{Companion Proceedings of
  the 36th International Conference on Software Engineering}}. ACM,
  \bibinfo{pages}{727--729}.
\newblock


\bibitem[\protect\citeauthoryear{Souza, Borges, d?Amorim, and
  P{\u{a}}s{\u{a}}reanu}{Souza et~al\mbox{.}}{2011}]%
        {souza2011coral}
\bibfield{author}{\bibinfo{person}{Matheus Souza}, \bibinfo{person}{Mateus
  Borges}, \bibinfo{person}{Marcelo d?Amorim}, {and} \bibinfo{person}{Corina~S
  P{\u{a}}s{\u{a}}reanu}.} \bibinfo{year}{2011}\natexlab{}.
\newblock \showarticletitle{CORAL: solving complex constraints for symbolic
  pathfinder}. In \bibinfo{booktitle}{\emph{NASA Formal Methods Symposium}}.
  Springer, \bibinfo{pages}{359--374}.
\newblock


\bibitem[\protect\citeauthoryear{Trinh, Chu, and Jaffar}{Trinh
  et~al\mbox{.}}{2014}]%
        {trinh2014s3}
\bibfield{author}{\bibinfo{person}{Minh-Thai Trinh}, \bibinfo{person}{Duc-Hiep
  Chu}, {and} \bibinfo{person}{Joxan Jaffar}.} \bibinfo{year}{2014}\natexlab{}.
\newblock \showarticletitle{S3: A symbolic string solver for vulnerability
  detection in web applications}. In \bibinfo{booktitle}{\emph{Proceedings of
  the 2014 ACM SIGSAC Conference on Computer and Communications Security}}.
  ACM, \bibinfo{pages}{1232--1243}.
\newblock


\bibitem[\protect\citeauthoryear{Veanes, Bj{\o}rner, and De~Moura}{Veanes
  et~al\mbox{.}}{2010}]%
        {veanes2010symbolic}
\bibfield{author}{\bibinfo{person}{Margus Veanes}, \bibinfo{person}{Nikolaj
  Bj{\o}rner}, {and} \bibinfo{person}{Leonardo De~Moura}.}
  \bibinfo{year}{2010}\natexlab{}.
\newblock \showarticletitle{Symbolic automata constraint solving}. In
  \bibinfo{booktitle}{\emph{International Conference on Logic for Programming
  Artificial Intelligence and Reasoning}}. Springer, \bibinfo{pages}{640--654}.
\newblock


\bibitem[\protect\citeauthoryear{Visser, Geldenhuys, and Dwyer}{Visser
  et~al\mbox{.}}{2012}]%
        {visser2012green}
\bibfield{author}{\bibinfo{person}{Willem Visser}, \bibinfo{person}{Jaco
  Geldenhuys}, {and} \bibinfo{person}{Matthew~B Dwyer}.}
  \bibinfo{year}{2012}\natexlab{}.
\newblock \showarticletitle{Green: reducing, reusing and recycling constraints
  in program analysis}. In \bibinfo{booktitle}{\emph{Proceedings of the ACM
  SIGSOFT 20th International Symposium on the Foundations of Software
  Engineering}}. ACM, \bibinfo{pages}{58}.
\newblock


\bibitem[\protect\citeauthoryear{Visser, P?s?reanu, and Khurshid}{Visser
  et~al\mbox{.}}{2004}]%
        {visser2004test}
\bibfield{author}{\bibinfo{person}{Willem Visser}, \bibinfo{person}{Corina~S
  P?s?reanu}, {and} \bibinfo{person}{Sarfraz Khurshid}.}
  \bibinfo{year}{2004}\natexlab{}.
\newblock \showarticletitle{Test input generation with Java PathFinder}.
\newblock \bibinfo{journal}{\emph{ACM SIGSOFT Software Engineering Notes}}
  \bibinfo{volume}{29}, \bibinfo{number}{4} (\bibinfo{year}{2004}),
  \bibinfo{pages}{97--107}.
\newblock


\bibitem[\protect\citeauthoryear{Wagner, Foster, Brewer, and Aiken}{Wagner
  et~al\mbox{.}}{2000}]%
        {wagner2000first}
\bibfield{author}{\bibinfo{person}{David Wagner}, \bibinfo{person}{Jeffrey~S
  Foster}, \bibinfo{person}{Eric~A Brewer}, {and} \bibinfo{person}{Alexander
  Aiken}.} \bibinfo{year}{2000}\natexlab{}.
\newblock \showarticletitle{A First Step Towards Automated Detection of Buffer
  Overrun Vulnerabilities.}. In \bibinfo{booktitle}{\emph{NDSS}}.
  \bibinfo{pages}{2000--02}.
\newblock


\bibitem[\protect\citeauthoryear{Wagner, Kuznetsov, and Candea}{Wagner
  et~al\mbox{.}}{2013}]%
        {wagner2013overify}
\bibfield{author}{\bibinfo{person}{Jonas Wagner}, \bibinfo{person}{Volodymyr
  Kuznetsov}, {and} \bibinfo{person}{George Candea}.}
  \bibinfo{year}{2013}\natexlab{}.
\newblock \showarticletitle{-Overify: Optimizing Programs for Fast
  Verification}. In \bibinfo{booktitle}{\emph{14th Workshop on Hot Topics in
  Operating Systems (HotOS XIV)}}.
\newblock


\bibitem[\protect\citeauthoryear{Yang, P{\u{a}}s{\u{a}}reanu, and
  Khurshid}{Yang et~al\mbox{.}}{2012}]%
        {yang2012memoized}
\bibfield{author}{\bibinfo{person}{Guowei Yang}, \bibinfo{person}{Corina~S
  P{\u{a}}s{\u{a}}reanu}, {and} \bibinfo{person}{Sarfraz Khurshid}.}
  \bibinfo{year}{2012}\natexlab{}.
\newblock \showarticletitle{Memoized symbolic execution}. In
  \bibinfo{booktitle}{\emph{Proceedings of the 2012 International Symposium on
  Software Testing and Analysis}}. ACM, \bibinfo{pages}{144--154}.
\newblock


\bibitem[\protect\citeauthoryear{Yu, Bultan, and Ibarra}{Yu
  et~al\mbox{.}}{2009}]%
        {yu2009symbolic}
\bibfield{author}{\bibinfo{person}{Fang Yu}, \bibinfo{person}{Tevfik Bultan},
  {and} \bibinfo{person}{Oscar~H Ibarra}.} \bibinfo{year}{2009}\natexlab{}.
\newblock \showarticletitle{Symbolic string verification: Combining string
  analysis and size analysis}. In \bibinfo{booktitle}{\emph{International
  Conference on Tools and Algorithms for the Construction and Analysis of
  Systems}}. Springer, \bibinfo{pages}{322--336}.
\newblock


\bibitem[\protect\citeauthoryear{Zheng, Zhang, and Ganesh}{Zheng
  et~al\mbox{.}}{2013}]%
        {zheng2013z3}
\bibfield{author}{\bibinfo{person}{Yunhui Zheng}, \bibinfo{person}{Xiangyu
  Zhang}, {and} \bibinfo{person}{Vijay Ganesh}.}
  \bibinfo{year}{2013}\natexlab{}.
\newblock \showarticletitle{Z3-str: A z3-based string solver for web
  application analysis}. In \bibinfo{booktitle}{\emph{Proceedings of the 2013
  9th Joint Meeting on Foundations of Software Engineering}}. ACM,
  \bibinfo{pages}{114--124}.
\newblock


\end{thebibliography}

\end{document}